\documentclass[fleqn,letterpaper,11pt]{scrartcl}

\usepackage{times}

\usepackage[section] {placeins}
\usepackage{graphicx}
\usepackage{amsmath}
\usepackage{fancyhdr}
\usepackage{enumerate}
\usepackage{indentfirst}
\usepackage{xspace}
\usepackage{array}
\usepackage{rotating}
\usepackage{bbold}
\usepackage{cite}
\usepackage[
  pdfusetitle,
  pdfkeywords={minerva},
  bookmarks,bookmarksopen,
  bookmarksnumbered,
  pdfpagelabels,
  colorlinks,linkcolor=black,citecolor=black,
  pdfview={Fit},
]
{hyperref}

\newif\ifpdf
\ifx\pdfoutput\undefined
   \pdffalse
\else
   \pdfoutput=1
   \pdftrue
\fi
\ifpdf
   \usepackage{graphicx}
   \usepackage{epstopdf}
\else
  \usepackage{graphicx}
\fi

\newcommand{\minerva}{MINER$\nu$\kern-0.12emA\xspace}

\begin{document}  

.

\vspace{2ex}
\begin{center} 
{\LARGE GENIE Implementation of IFIC Valencia \\QE-like 2p2h cross section } \\
\vspace{2ex}
{\large Jackie Schwehr, Dan Cherdack, T2K experiment, Colorado State University,\\
Rik Gran, MINERvA experiment, University of Minnesota Duluth} \\
\vspace{2ex}
\today
\end{center}

\begin{abstract}

The model by Nieves, Ruiz-Simo, Vicente Vacas, and their group (IFIC, Valencia, Spain)
for 2p2h reactions that
produce QE-like (no pion) final states has been implemented in GENIE.
Since the model currently does not predict the kinematics of the
outgoing hadrons, a simple two-nucleon system is grafted onto the model's
prediction of isospin, energy transfer, and momentum transfer.  These two
nucleons are then given to the GENIE FSI models.
This technical note is a guide to the
kind of information available from this model and some limitations.
There are several figures that illustrate the output of the model, and
detailed discussion of the physics context for this model.
Finally, any other authors' model (or variations of this one) that can
be expressed as hadronic tensors for total and pn initial state will
be easy to incorporate into this framework, or possibly be made
available as a reweight to events generated with this model.  
The 2017 version of this
document is updated to reflect the as-released GENIE 2.12.6 version of
the code, which produces identical results to the development versions.

\end{abstract}

\pagebreak

\tableofcontents

\pagebreak

\section{Introduction}

This technical note describes the initial GENIE
implementation of the Valencia group's
model for QE-like two-particle, two-hole (2p2h) (or meson exchange
current MEC) reactions.  
This is one effort in a
longstanding campaign by members of the Instituto de Fisica
Corpuscular (IFIC) to describe neutrino nucleus reactions.   Now available
in GENIE, their work is available for wider use.
We will
refer to these as ``IFIC Valencia model''.  It is sometimes referred to as the
Nieves model or the Valencia model, and the primary authors are
Nieves, Ruiz Simo, and Vicente Vacas with contributions from Oset and
Alvarez Ruso at IFIC.
The model is described in references
\cite{Nieves:2011pp, Gran:2013kda}.  This implementation incorporates a
kinematic cutoff at 1.2 GeV of momentum transfer, 
described in the latter reference, allowing it to be used at higher energies.  

With a QE model that importantly includes the RPA
effect and a local Fermi-gas \cite{Nieves:2004wx}, the combination describes MiniBooNE QE-like p$_\mu$,
$\theta_\mu$ data \cite{Nieves:2011yp} and MINERvA QE-like $Q^2$
cross section shape \cite{Gran:2013kda} better than either
experiments' default QE and Delta background models.
This implementation, available as of version 2.12.6 of GENIE, was used
(in the form of a back-ported private build of version 2.8.4) for the
comparisons in \cite{Rodrigues:2015hik} and later MINERvA papers,
for which the present paper serves as a technical reference.
When combined with updated deuterium form factor constraints such as
\cite{Meyer:2016oeg} and uncertainties on the RPA effect
\cite{Gran:2017psn} for example,
these elements can replace the anomalously high axial
mass and its uncertainty.   This enables the community to proceed with
precision measurements of these processes.

The cross section calculation is fast because the hadronic
tensor is pre-computed and available in tabulated form.  
Essentially all the physics is related to the energy and momentum transfer from
the lepton to the nucleus.  These tensor tables
serve for all energies, for neutrino and anti-neutrino, and all lepton flavors.  
It is contracted with the leptonic tensor, a fast calculation with no integrations, which account
for the lepton mass and energy.  That this would be a productive
implementation originated with Federico Sanchez as part of work on the
\cite{Gran:2013kda} paper; Sanchez also wrote the first C++ skeleton
code that calls the model authors' {\small FORTRAN} routines.   The GENIE
code is a combination of this skeleton and the previous MEC model in GENIE.

There is a tensor table each for the full cross section for three nuclei,
$^{12}$C, $^{16}$O, and $^{40}$Ca from the original references
\cite{Nieves:2011yp,Gran:2013kda}, plus four additional nuclei.  These
are enough to cover the rest of the periodic table with a nucleon
combinatorics scaling, described in Sec.~\ref{sec:othernuclei}.
Additional hadronic tensor 
tables provide the cross section for scattering on pn initial nucleon 
states, which are used internally to generate the isospin final states 
according to the model.  Two more tensor tables for each nuclei are provided for a
$\Delta$-only version of the full and pn cross section;  the
physics of this component is discussed in Sec.~\ref{sec:delta}, and
may be especially useful for systematics studies.

The hadron tensor part of the technique described here is not specific to these authors'
model.
Other model authors' work, or future variations on the IFIC model, can be dropped into
GENIE simply by providing a revised hadronic tensor.  For example, the
earliest full calculation of a neutrino-carbon 2p2h model was done by Martini and Ericson
\cite{Martini:2009uj} and could be available within GENIE using this
same codebase.   The technique may even be useful to make fast comparisons of
different author's QE process.
Some comments on this are available in a later section.

Mosts figures in this document are done with pre-release version
and GENIE 2.8 and 2.10 and all default settings.  The results are
the same when obtained from the released version GENIE 2.12.6 .  The
hadronic tensors used are from the original author's version HP4.2
which was used for most of the 2013-era paper \cite{Gran:2013kda}.  
We are using a notation in this document where $q_0$ is the energy
transfer, sometimes called $\omega$ or $\nu$, and $q_3$ is the
magnitude of the three-momentum transfer, also naturally called
$|\vec{p}|$ or sometimes just $p$.  
These are the components of a four vector such that $-q^2 = Q^2 =
q_3^2 - q_0^2$ is the invariant square of the four-momentum transfer.

\begin{figure}[htbp]
\begin{center}
\includegraphics[width=7.0cm]{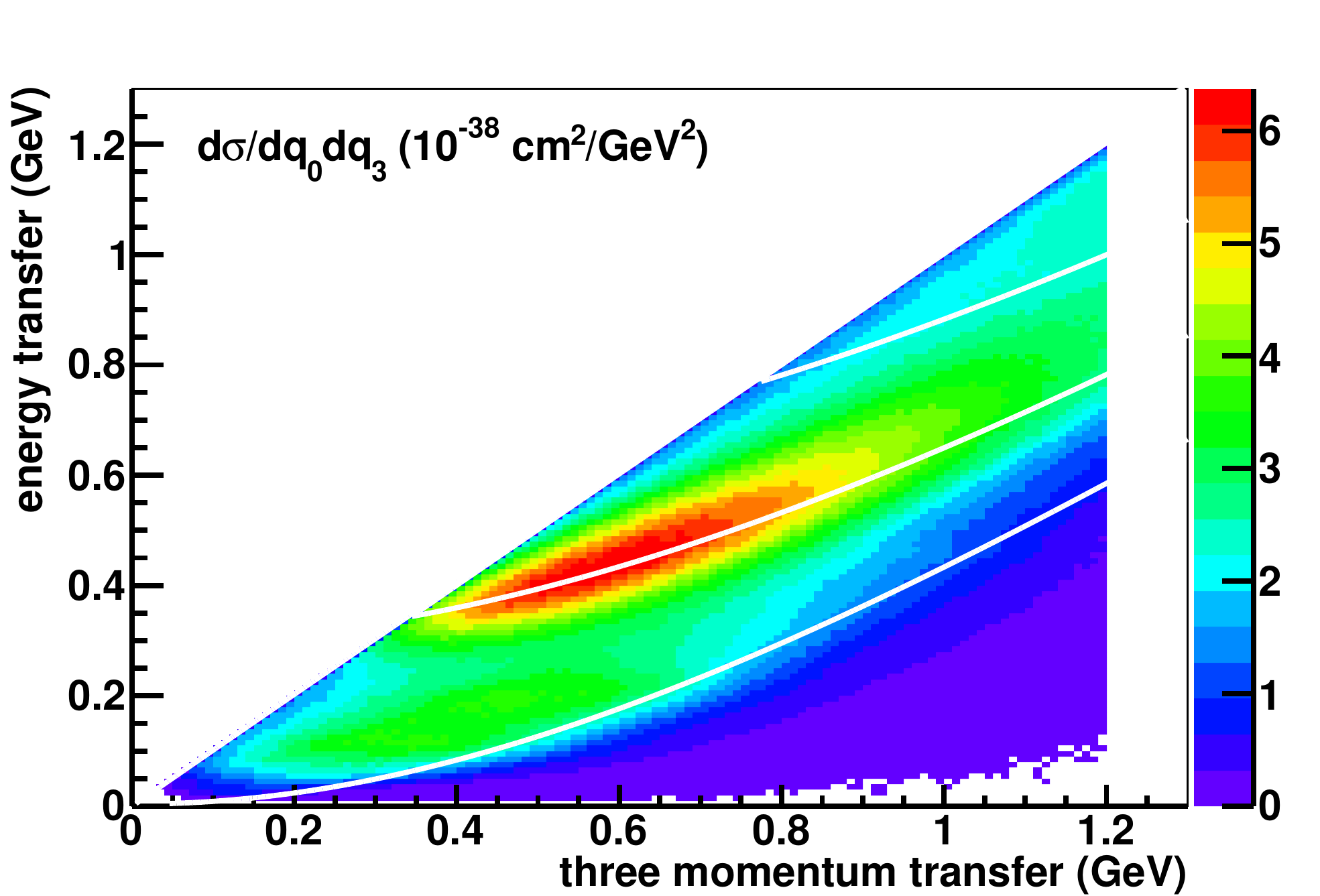}
\includegraphics[width=7.0cm]{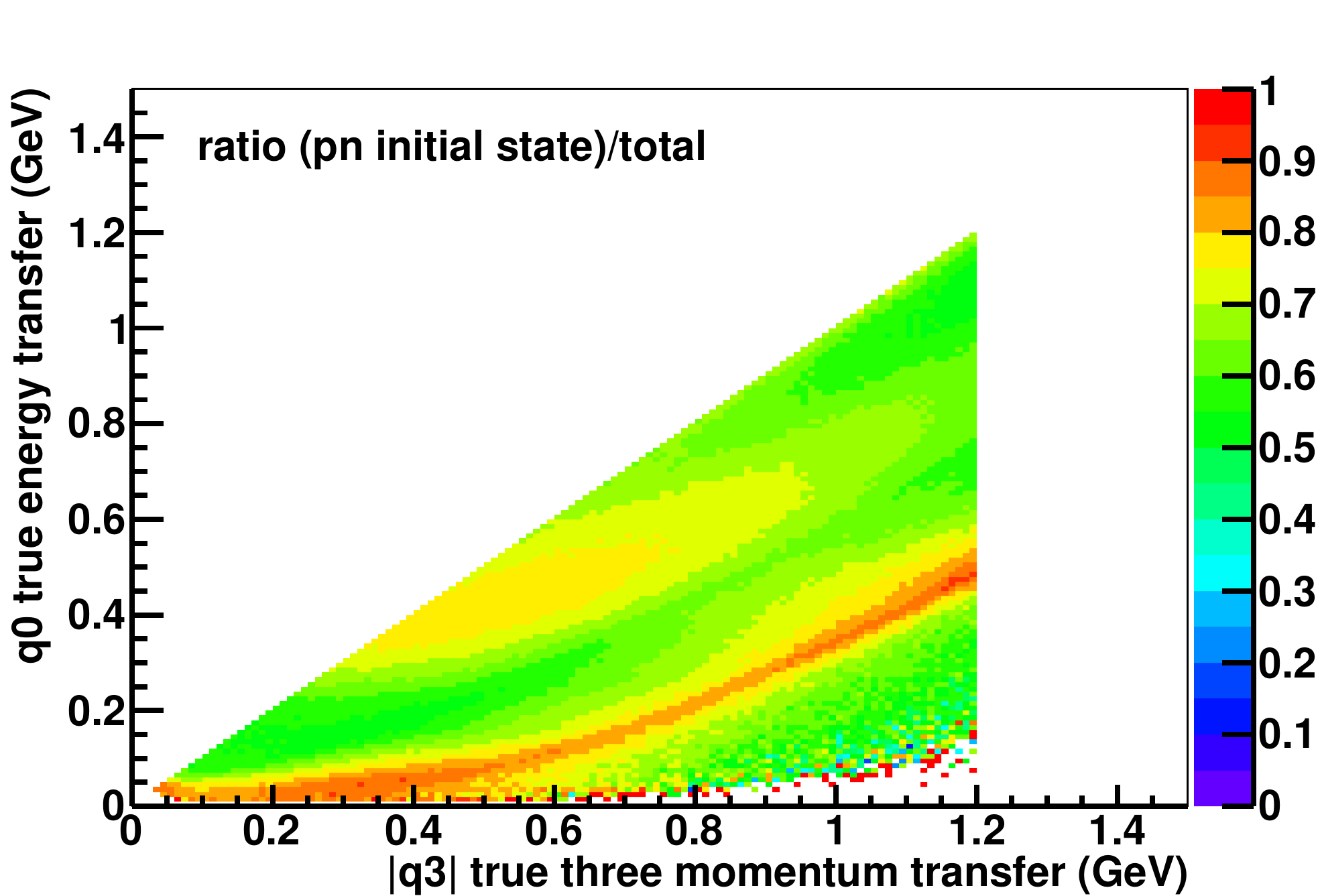}
\includegraphics[width=7.0cm]{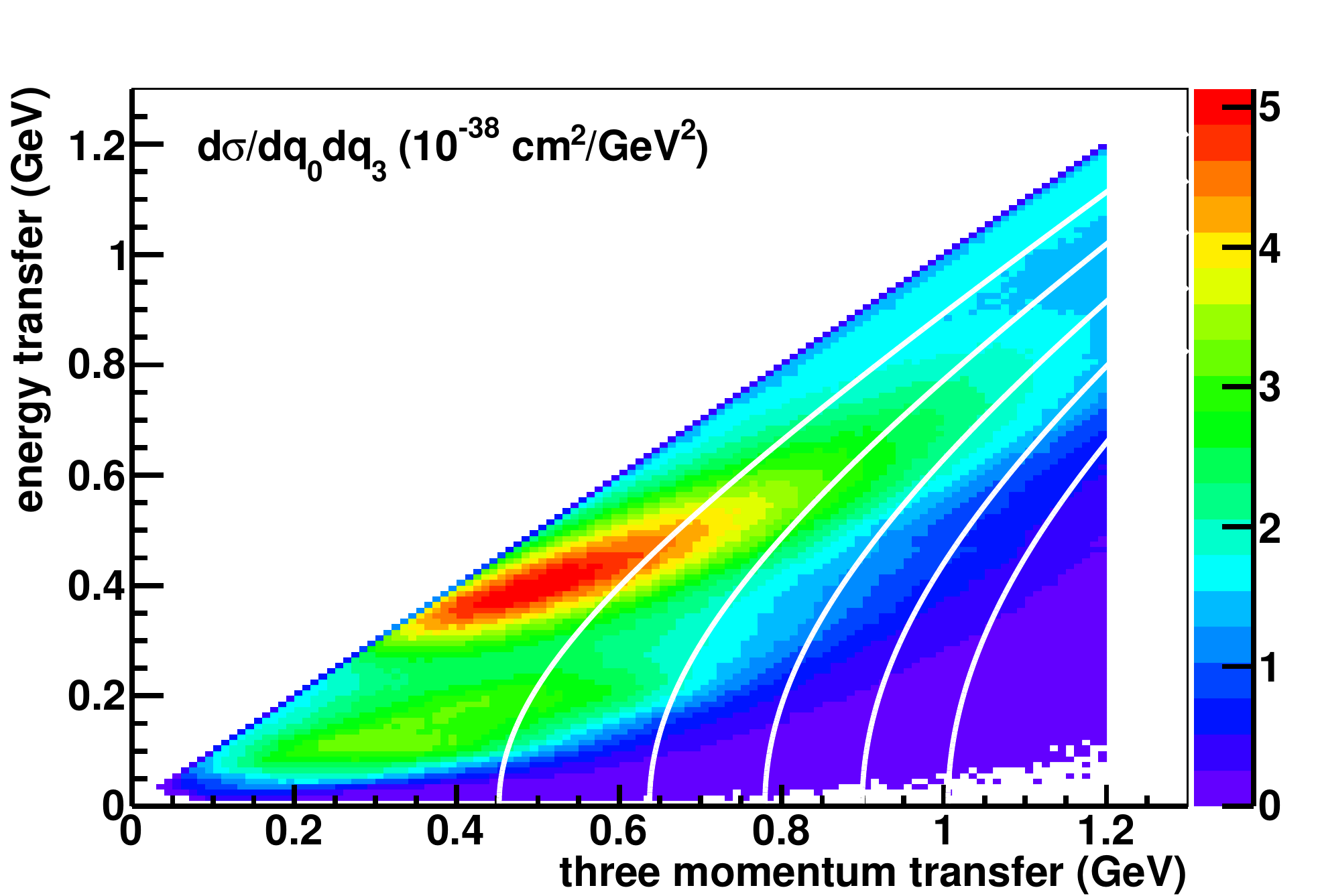}
\includegraphics[width=7.0cm]{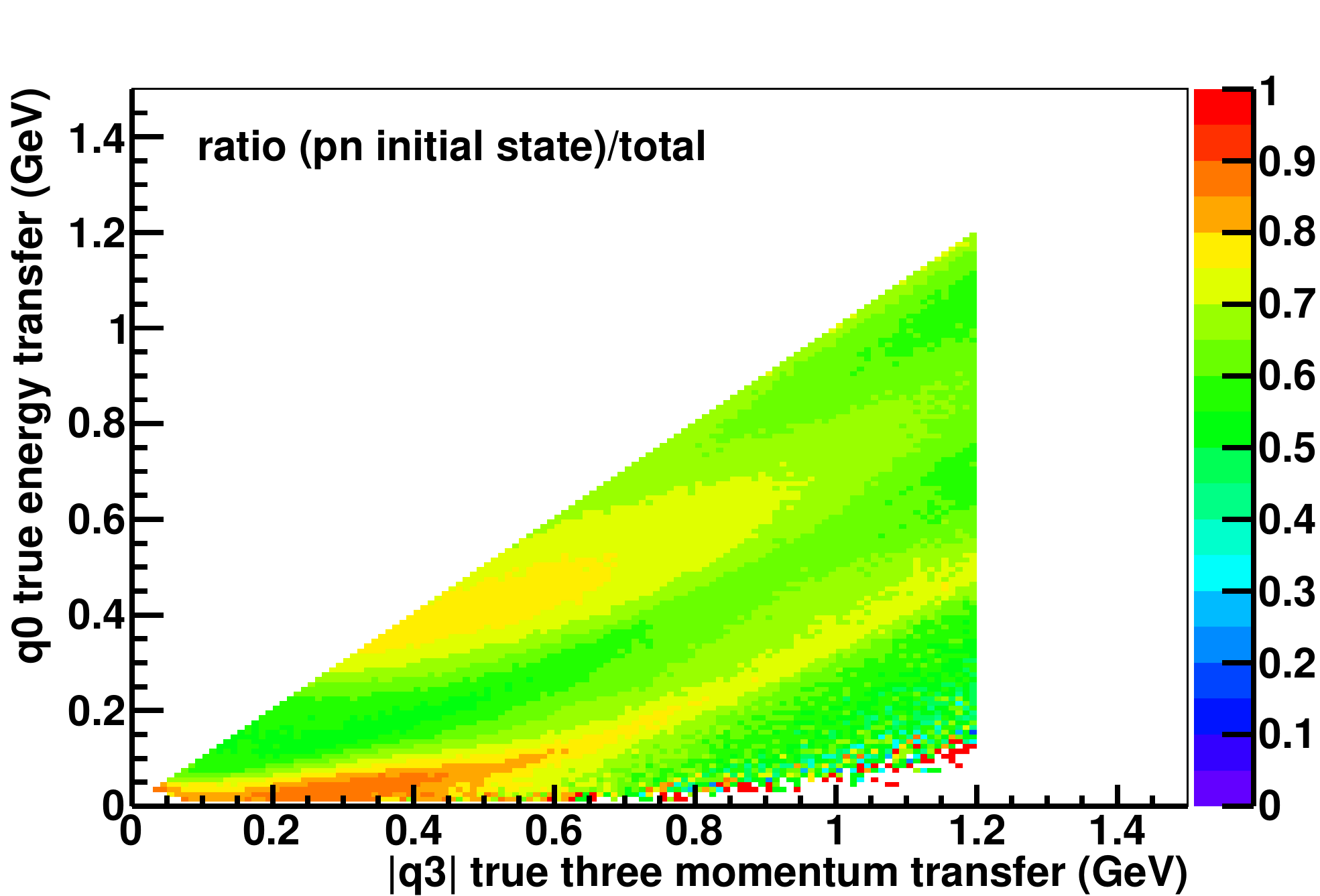}

\parbox{6in}{\caption{The differential cross section from the GENIE
    implementation of the QE-like 2p2h model (right) and the fraction
    of the total cross section with a pn initial state (left).  The
    top plots are
    $\nu$C while the lower plots are $\bar{\nu}$C, both at 3 GeV.
  To guide the eye in this kinematics space, the neutrino figure has
  lines of constant W = 938, 1232, 1520 MeV emphasizing the dip region, and the anti-neutrino
  figure has lines of constant $Q^2$ from 0.2 to 1.0 GeV$^2$
  emphasizing the low $Q^2$ nature of the cross section.
\label{fig:diffxs}}}
\end{center}
\end{figure}

Cross sections are often calculated for muon kinematics p$_\mu$ and
$\theta_\mu$ in the lab/nucleus rest frame.  Here we present it in the
energy and momentum transfer quantities.  Figure~\ref{fig:diffxs} shows
the cross section overlaid with lines of constant W = 0.938, 1.232,
and 1.535~GeV.
The resulting QE-like 2p2h differential cross section has two components, a
$\Delta$(1232)-like component and a
dip-region component just above W=0.938~GeV.  
The lower plots are the anti-neutrino cross section, but with lines of constant $Q^2$.
The cross section shown is obtained from 3 GeV neutrinos by generating
millions of events on
carbon nuclei, binning them, and scaling by the total $\sigma$
captured from the model available in the GENIE spline value.
The right plots are the model's prediction for the fraction of
events that had pn initial states, which will be pp final states in
the neutrino case (top) and nn final states for anti-neutrinos (bottom).  This is the essential information available
from the model.  These predictions change only slightly switching from
muon to the lighter electron. 

After the lepton kinematics are chosen for each generated event, the
GENIE implementation grafts a hadron system final state.
This discards some correlation of the
cross section with the initial nucleon system, and simplifies how the
two nucleons share the energy and momentum transfer.  The full
calculation is not available in a hadronic tensor style
implementation, because the detailed hadron kinematics are hidden in the inner
integrations of the model, so this is the best we can do.  This
initial implementation, combined with data from recent experiments,
may promote future work on the details of the model.  This hadron methodology
is similar in design to what is implemented \cite{Sobczyk:2012ms} in NuWro and
in the phenomenological MEC model in GENIE authored by Steve Dytman.
This simple method appears to be a good approximation
\cite{Simo:2014esa}
to the proton angles from a full calculation.

A beneficial consequence of building the hadron state independent from
(except $q_0$,$q_3$) the lepton state is that an event generated with
the IFIC Valencia model in GENIE is necessarily the same as would be
generated from an alternate model which only provides
$d^2\sigma/dq_0q_3$ and not correlated hadron information.  That permits
opportunities to reweight events generated with one model to be
identical to a variation of (or another author's) model for
$d^2\sigma/dq_0q_3$.  For the same reason, it is possible to reweight a
quasi-elastic model to another $d^2\sigma/dq_0q_3$, an approximation that misses only some
correlation expected between the initial and final nucleon state that
may be present in one model or the other.  Especially when the final
nucleon states are further modified by rescattering as they exit the
nucleus, the correlations may anyway be challenging to observe.
This strategy is widely exploited by the
experimental community to make fast comparisons or fitting or rapid
evaluation of systematic uncertainties. 

This paper outlines the basic procedural steps in the next section.
After that are a series of subsections that touch on specific outcomes
of this implementation, especially the resulting hadronic systems and 
where they might diverge from the model author's calculations.
Practical and optimization concerns mean that the resulting cross
section will not be identical to the original calculation.  However
all results are benchmarked against the original code, and the design
goal is to reproduce that with 2\% accuracy in the cross section
lepton kinematics.  For a prediction, this seems a good match for
testing against current and imminent data sets.

\section{Basic procedure}

\begin{itemize}
\item[] Use accept-reject method in $p_\mu$, $\theta_\mu$ 
  to draw momentum and energy transfer
  kinematics up to 1.2 GeV according to the IFIC Valencia model.
\item[] Assign the energy
  and momentum transfer to a pair of nucleons drawn from the user's
  favorite nucleon
  momentum distribution.
\item[] ``Decay'' that nucleon cluster back-to-back and isotropically
  to two nucleons in the center of momentum frame and boost back to the lab frame.
\item[] the two nucleons are available to be rescattered by the user's
  favorite final state interaction model.
\end{itemize}

The accept-reject method is seeded with a maximum cross section to
improve generation efficiency.  
Even though the cross section is approximately the same as a function
of $q_0$ and $q_3$,  the method is applied in the not-invariant kinematics $p_\mu$
and $\theta_\mu$.  This means that the maximum double-differential cross section for any
muon kinematics increases with energy.   A simple but safe log-log
parameterization was created once and hard-coded, then used to compute
a maximum cross section.   (This technical issue could be replaced by a
fast lookup of a cached value.)  This maximum value
is scaled away from A=12 by a generous function A$^{1.4}$ because some
components may increases faster with A.  These are specific to
IFIC model, using Martini's model will require different
constants.  This does not change the physics of the model, just allows
more efficient generation of events.

Are the initial two nucleons pn or not ?  This is drawn from the cross
section also.   The hadron tensor computed for the entire
cross section was used when the kinematics were chosen.  A second
cross section is computed for
these same kinematics for the pn initial state only.
The ratio of the pn cross section is taken with the total cross section.  Then a random number
is compared to that ratio to decide whether the pn initial state is chosen. 


The initial nucleon pair is drawn from the prevailing GENIE Fermi
momentum distribution.  For Genie 2.12 and and earlier it is a Fermi gas with
the Bodek-Ritchie high momentum tail.  The four-momentum sum of the
nucleons is computed and saved to the GHEP record as a nucleon
cluster.  In this implementation, the original momentum
of each nucleon is forgotten.  If the user chooses a non-default
Fermi momentum distribution, that will be used.

The momentum and energy transfer, plus one unit
charge, is given to the nucleon pair.  
If the resulting nucleon pair would be off shell, then we loop and rethrow for
another nucleon pair. 

The nucleon pair is then ``decayed'' using a CERNLIB era phase space
decayer available through ROOT.
Again, this process knows nothing about the initial momenta of
the nucleons, only the total four-momentum of the system.  The
``decay'' is back to back and isotropic in the center of momentum system, and is then
boosted back to the lab/nucleus rest frame.

The resulting nucleons are available to be rescattered using the
prevailing FSI model in GENIE.  There is a very high probability of at
least a soft, elastic scatter, and better than 50\% chance that three
or more nucleons will exist the nucleus.

The code also has a method which will return the integrated cross section,
for making splines and other uses.  It is simply a numerical
integration of the cross section computed in $q_0$,$q_3$ kinematic
space.  
The GENIE procedure is to produce a spline with some points that span
an energy range logarithmically, then use a cubic spline interpolation to get a
specific energy within the tabulated points.  For the 3 GeV neutrino -
carbon reaction, it returns 1.4643 x 10$^{-38}$ cm$^2$, with 1.0943
x 10$^{-38}$ cm$^2$ for anti-neutrino .  A high resolution integration
directly from the model yields 1.4621 x 10$^{-38}$, which is the
same.  

The cross sections for electron neutrino interactions are slightly
higher, because the lower lepton mass opens up slightly more phase
space.  It especially appears along the line of maximum $q_0$ and
$q_3$,  all of which is kinematically at $Q^2\sim0$~GeV$^2$.
It is 1.3\% different at 3 GeV (1.5\% for anti-neutrinos), and about
6\% higher at 1 GeV (7\% different for anti-neutrinos) compared to the
prediction for muon neutrinos.

\begin{figure}[htbp]
\begin{center}
\includegraphics[width=9.0cm]{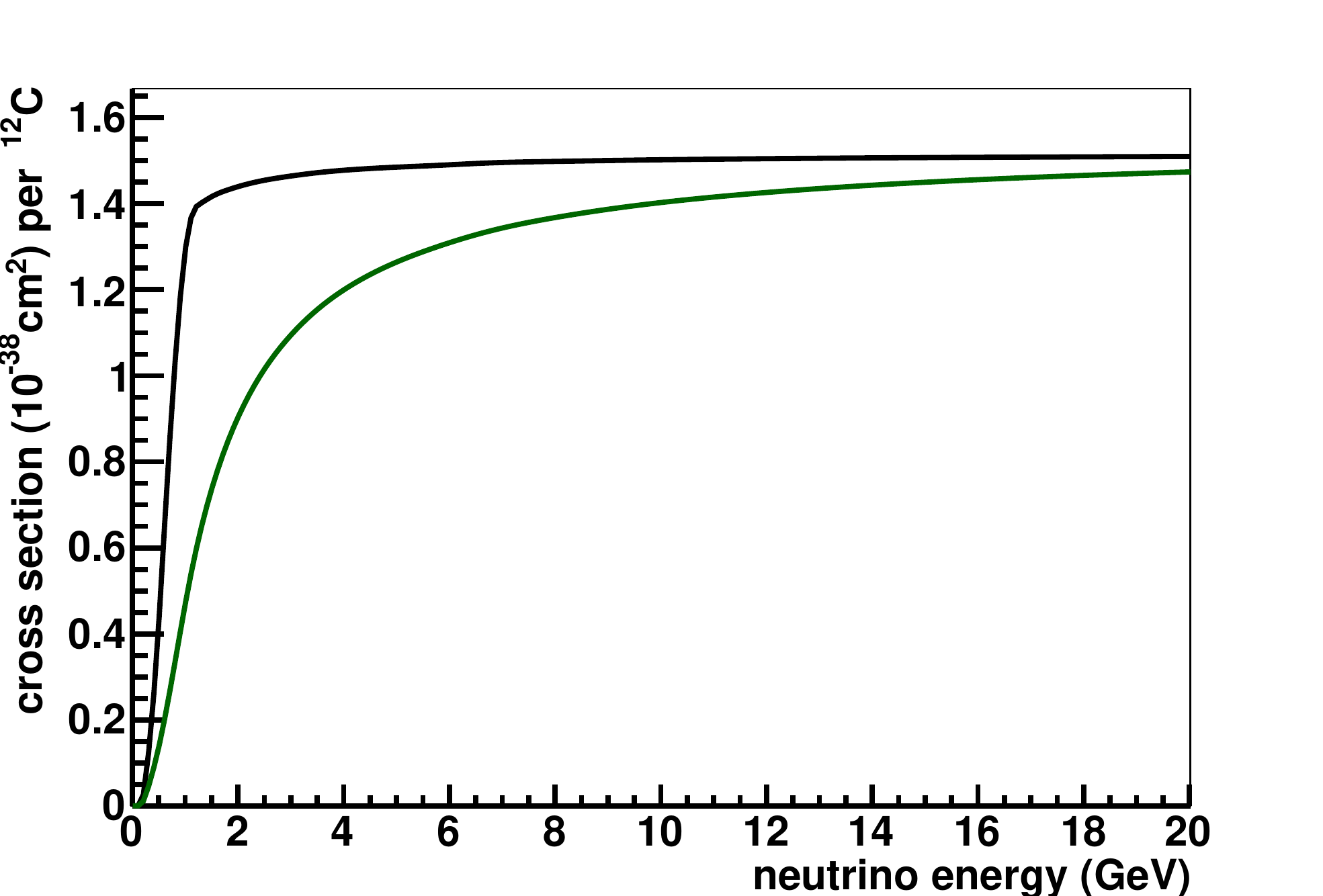}

\parbox{6in}{\caption{The $\sigma(E)$ cross section from the GENIE
    implementation of the QE-like 2p2h model, including the kinematic
    cutoff $q_3 < 1.2$ GeV.  This is the neutrino
    (black) and anti-neutrino (green) spline.
\label{fig:splines}}}
\end{center}
\end{figure}

A plot of the nu-C (black) and nubar-C (green) $\sigma(E)$ 
splines generated by GENIE are shown in Fig.~\ref{fig:splines}.  The
sharp behavior in the neutrino case happens as the neutrino energy
drops below 1.2 GeV and eats into the limited kinematic space of this
version of the model.  The anti-neutrino case has a slower turn-on of
the cross section because of the axial interference term.  Both these
characteristics follow the familiar trends of the QE cross
section.

\subsection*{Q2 and Enu distributions}

\begin{figure}[htbp]
\begin{center}
\includegraphics[width=7.0cm]{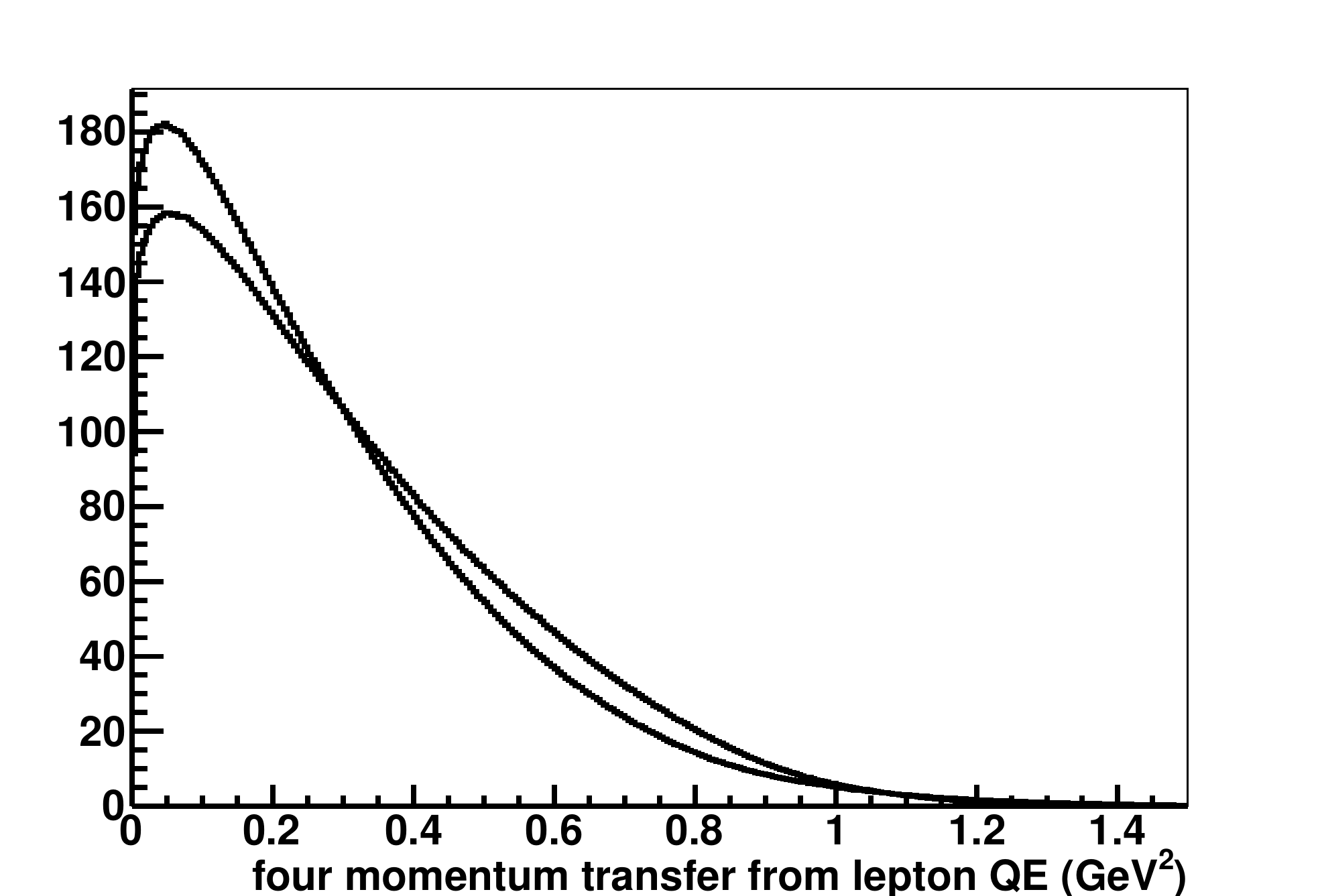}
\includegraphics[width=7.0cm]{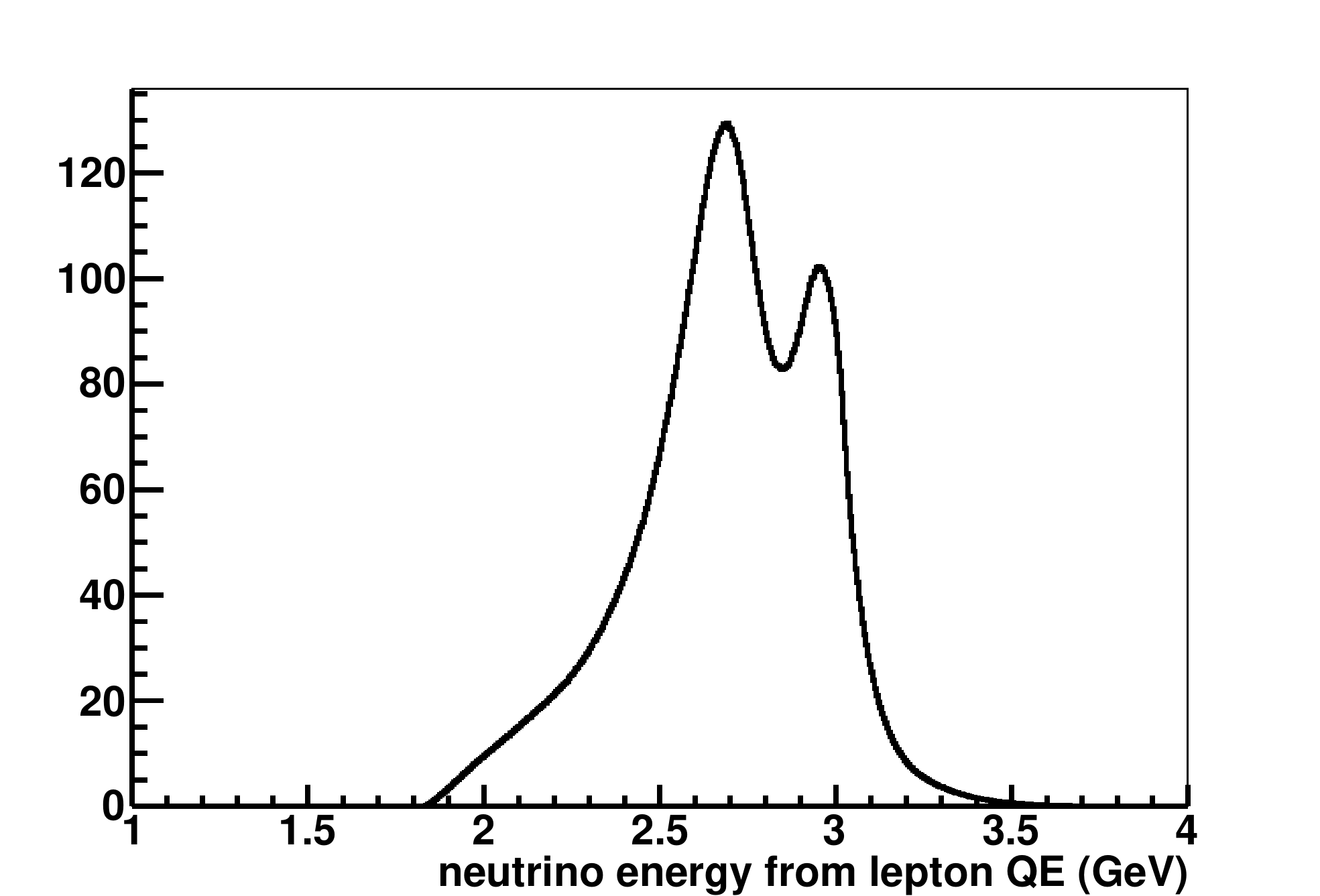}

\parbox{6in}{\caption{The model's $Q^2$ distribution (flatter), and one with
   the QE hypothesis (more peaked at zero) on the left show the effect
   of using only the muon kinematics to reconstruct these events.
   Also shown is the $E_\nu$
    distribution with the QE hypothesis on the right, for 3 GeV
    incident neutrinos.
\label{fig:q2enu}}}
\end{center}
\end{figure}

Traditionally, the QE and resonance region have been described using
the invariant square of the four-momentum transfer $Q^2$.  
Analyzing {\em data} using only lepton
kinematics (and QE hypothesis) causes biased energy and $Q^2$ estimators
for the fraction of QE-like non-QE reactions from resonances and 2p2h
processes.   Illustration of these effects, generated from these 
samples, are shown in Fig.~\ref{fig:q2enu}.

These events are predicted to be intrinsically very low $Q^2$, and
especially so when reconstructed using the
QE hypothesis.  The published $Q^2$ distributions and
differential cross sections from data all want fewer low-$Q^2$
interactions.  Compared to that, the
addition of this process would seem to worsen the agreement of the data and
a simple Fermi-gas model.  However, there is a similar but opposite
story with adding RPA effects to the QE process, which reduces the
predicted event rate at very low $Q^2$ more than this MEC process adds
it.  The combination of the two improve agreement with the data in
ways that neither alone can accomplish, see for example
\cite{Gran:2013kda} and especially \cite{Rodrigues:2015hik}.  
Variations of the QE process
with RPA are available from several calculations and if necessary can be
implemented as a $q_0$, $q_3$ or $Q^2$ reweight \cite{Gran:2017psn}.

Another concern is how the QE hypothesis is used when determining the
energy reconstruction.  These effects have long been known (e.g.
\cite{Gran:2006jn})  to bias reconstructed kinematics of the  
$\Delta$ process, but many authors are pointing it out anew
\cite{Nieves:2012yz,Martini:2012uc,Mosel:2013fxa} and quantifying the  
distortions of precision neutrino oscillation spectra for the MEC  
process. 
This is more important for Cerenkov detectors
than for calorimetric detectors, and is most critical when the model
is missing a non-QE process completely.  
For particles which are further ``above the QE line'' in W or
$q_0$, the QE hypothesis causes the incident neutrino energy
to be systematically underestimated.  Because
the neutrino energy estimate is used to compute $Q^2$, it too is
underestimated.
The left most peak in the right plot of Fig.~\ref{fig:q2enu} 
is primarily from the Delta-like events.  Like 1p1h Delta
events, if they were reconstructed with the QE hypothesis, the 3 GeV
events would have their energy estimated about 10\% low.  The peak near 3
GeV is from the band that keeps closest to the QE line.  They are also
biased low, but not as much.

\subsection*{Selected kinematics saved in the GENIE record}

GENIE allows a model to save what ``selected kinematics'' were used at
the vertex.  This implementation sets the inelasticity y =
$q_0$/$E_\nu$
and $Q^2$ = $q_3^2 - q_0^2$,
the final state lepton four vector, and the hadron (two-nucleon)
combined four-vector after the reaction.  We put in the hadron
invariant mass W also, built from
the two nucleons that we put in the nucleus, though its not especially 
interesting, nor is it accurate that the model really selected
it.    The selected t and Bjorken x variables that GENIE is willing to
save are not used.

\begin{figure}[htbp]
\begin{center}
\includegraphics[width=9.0cm]{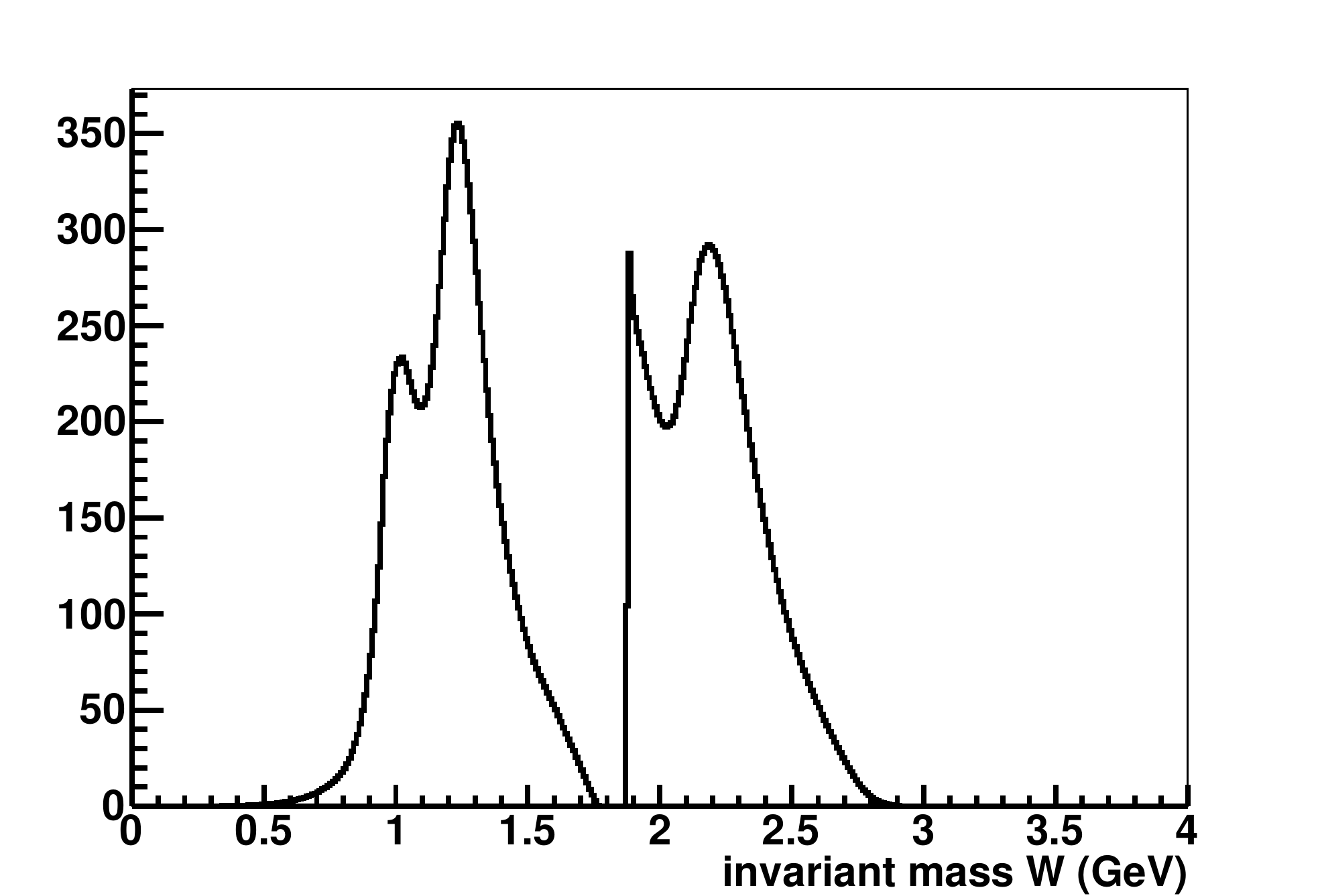}

\parbox{6in}{\caption{The two W distributions.  The right distribution
    is the actual W with two nucleons, before FSI.  The left distribution is the
    single-nucleon ``experimenter's W''.  The dip-region 2p2h population
    appears near W = 1.0 GeV in the latter, higher than QE events.
\label{fig:w}}}
\end{center}
\end{figure}

More interesting is the effective ``experimenter's W'', which
assumes a single nucleon $W^2 = M_n^2 + 2 M_n q_0 - Q^2$, which the
user can calculate later from either true or reconstructed quantities.
The proper two-nucleon W and the
experimenter's W are both shown in Fig.~\ref{fig:w}.  In each there
are two peaks, the right-most peak is the Delta-like component, and
the left-most peak is the more proper ``dip region'' component.

\section{Hadron kinematics distributions}

We emphasize the implementation of
the simplest hadron system and report what it produces.
How many nucleons are in the final state and what fraction goes to
protons are good observables for experiments that can do tracking
and/or calorimetry.
All the distributions below are with FSI, of course.  
Unlocking the information contained in the model might reveal
different energy sharing or angle dependence of the resulting
nucleons.  At this time we have not devised a way to provide an
uncertainty on this detail of the model; that may come with some operational
experience.

\subsection{nucleon multiplicity}

\begin{figure}[htbp]
\begin{center}
\includegraphics[width=7.0cm]{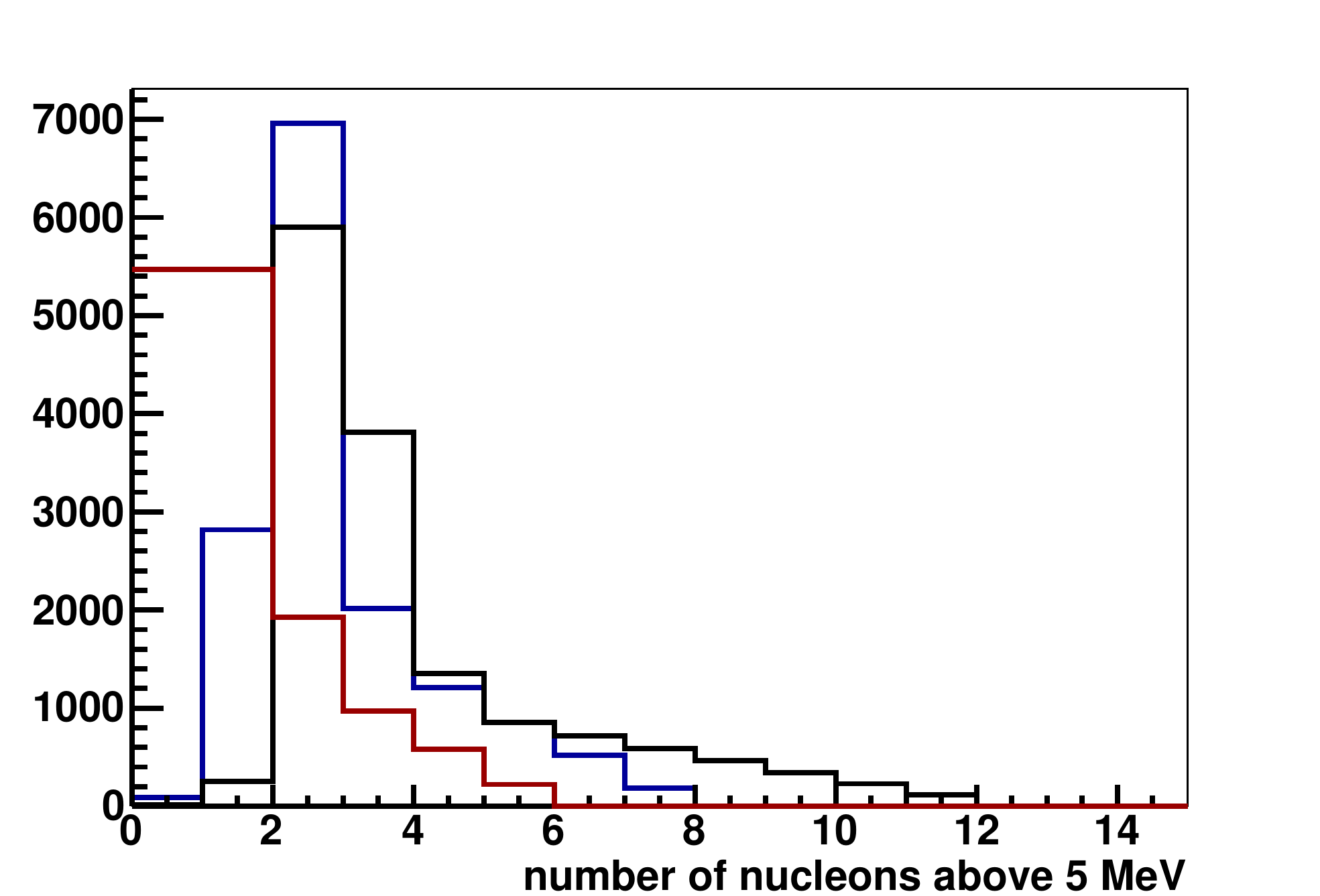}
\includegraphics[width=7.0cm]{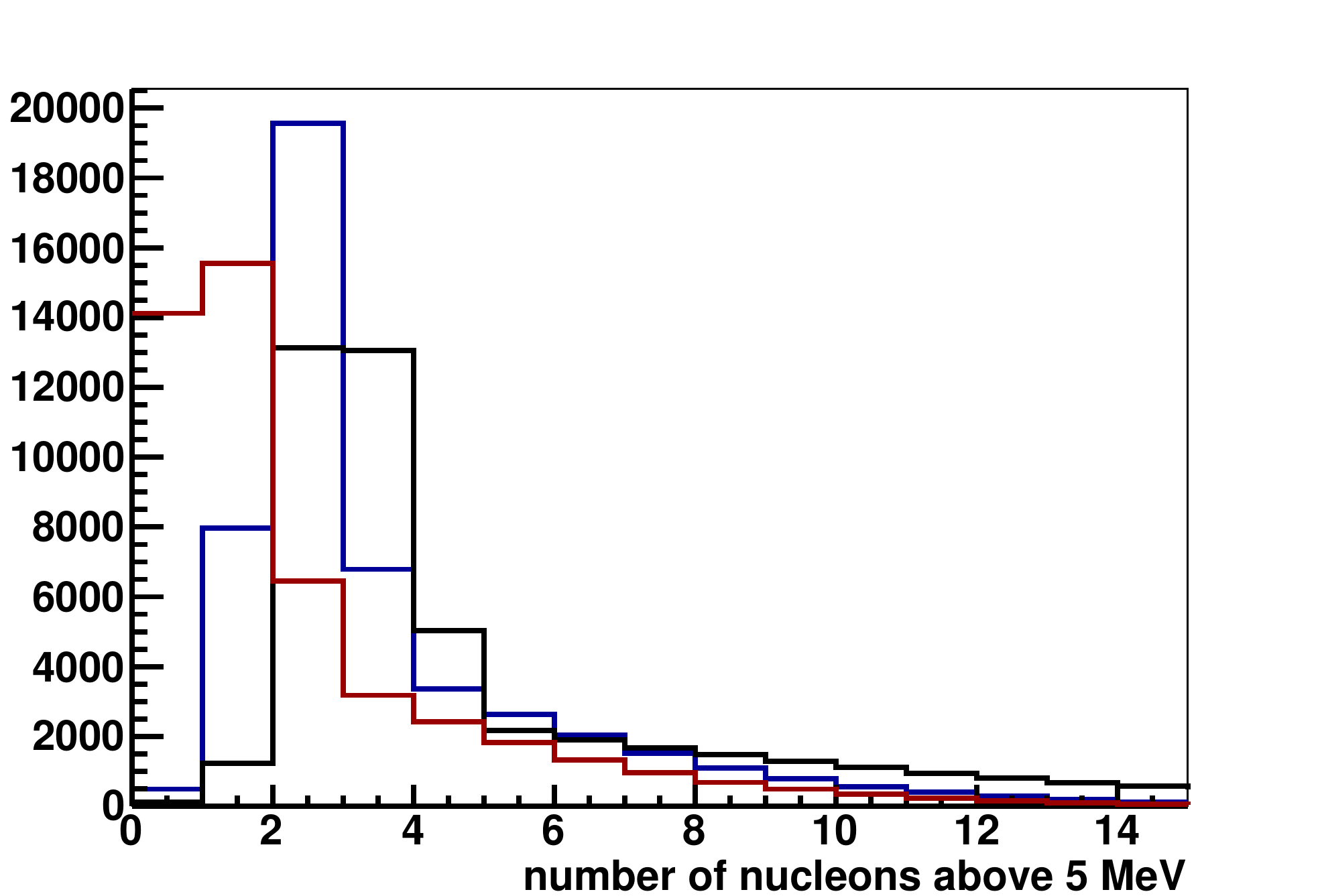}

\parbox{6in}{\caption{Number of nucleons (black), protons (blue) and
    neutrons (red) produced from two nucleons but after FSI.  For 3.0
    GeV neutrinos interacting in carbon-12 (left) and calcium-40 (right)
\label{fig:multiplicity}}}
\end{center}
\end{figure}

The interaction produces two nucleons specifically, and for $\nu$-C,
it prefers pp final states.  Even after intranuclear rescattering, it
almost never happens that there are zero protons (blue line), and the number of
neutrons is always less (red line), as shown in
Fig.~\ref{fig:multiplicity}.  The total nucleon multiplicity is shown
in black and starts at two nucleons as expected.  With reinteractions
via the GENIE FSI model, over half the events have three or more nucleons in
the final state.  The probability for nucleon FSI was already high, and
there are two nucleons making their way out of the nucleus in this
case.  The proton, neutron trends trade places for anti-neutrino
reactions, and the post-FSI GENIE final state will frequently have no protons in it.

\subsection{energy sharing}

\begin{figure}[htbp]
\begin{center}
\includegraphics[width=7.0cm]{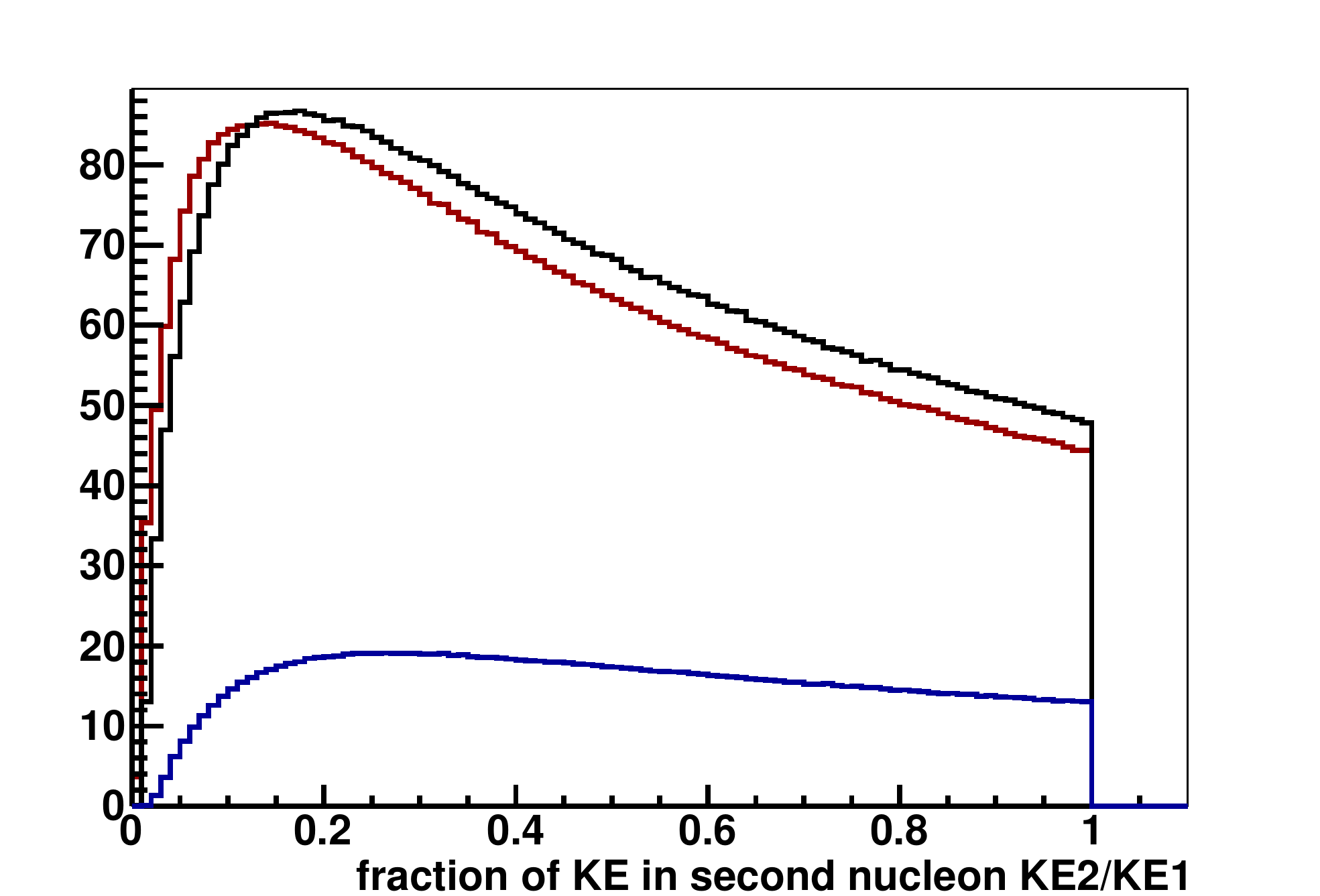}
\includegraphics[width=7.0cm]{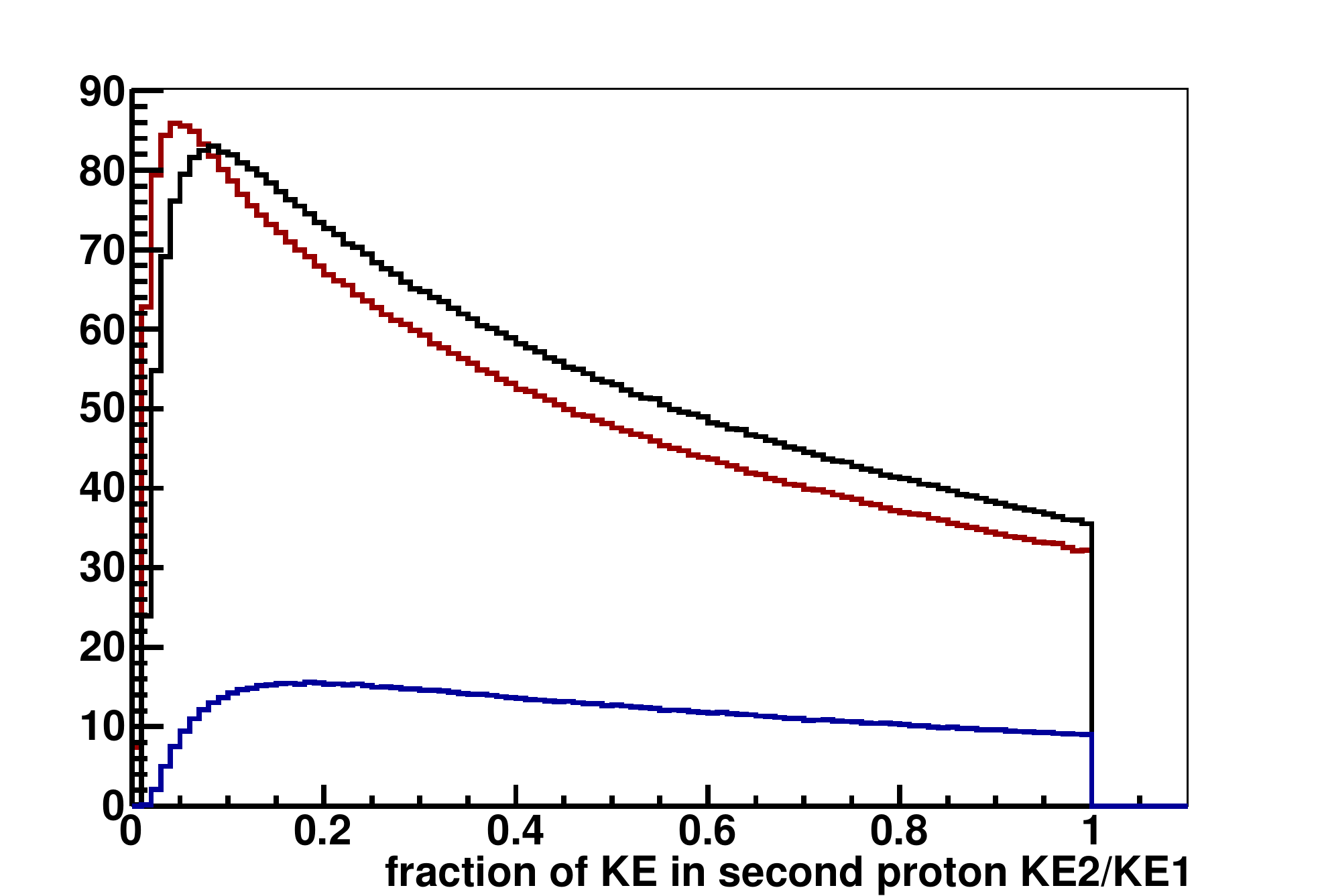}

\parbox{6in}{\caption{Energy sharing fraction (KE second nucleon)/(KE
    leading nucleon).  The left plot is for any of pp, pn, or nn, the
    right plot is for the most energetic two protons only. 
    Blue is $q_3<0.4$, black is $0.4 < q_3 < 0.8$ and red is $0.8 <
    q_3 < 1.2$~GeV sub samples.
\label{fig:fracsecond}}}
\end{center}
\end{figure}

The plot on the left in Fig.~\ref{fig:fracsecond}
illustrates how the nucleon energy is shared
after the isotropic decay is boosted to the lab frame and FSI is run.
The plot on the right is specifically for the most energetic two
protons.  Each figure has three lines representing the different
momentum transfer regions.
The nucleons are sorted by kinetic energy, and the horizontal axis is how much
fractionally is the second nucleon's KE is compared to the first.  The
distribution depends on how much boost the initial nucleon pair had
and so how asymmetric the products could be.  

At low $q_3$, the energy sharing (after the boost back to the lab frame
and after FSI)
spans the whole range of possibilities.  There is a good probability that the two nucleons
are sharing the energy equally, and since they are low energy, neither
travels very far.  In a coarse grained scintillator detector, they may
deposit signals in only one channel.  There are higher energy
transfers available at higher $q_3$, producing nucleons that
are much more asymmetric.



\subsection{available energy and missing energy}

\begin{figure}[htbp]
\begin{center}
\includegraphics[width=7.0cm]{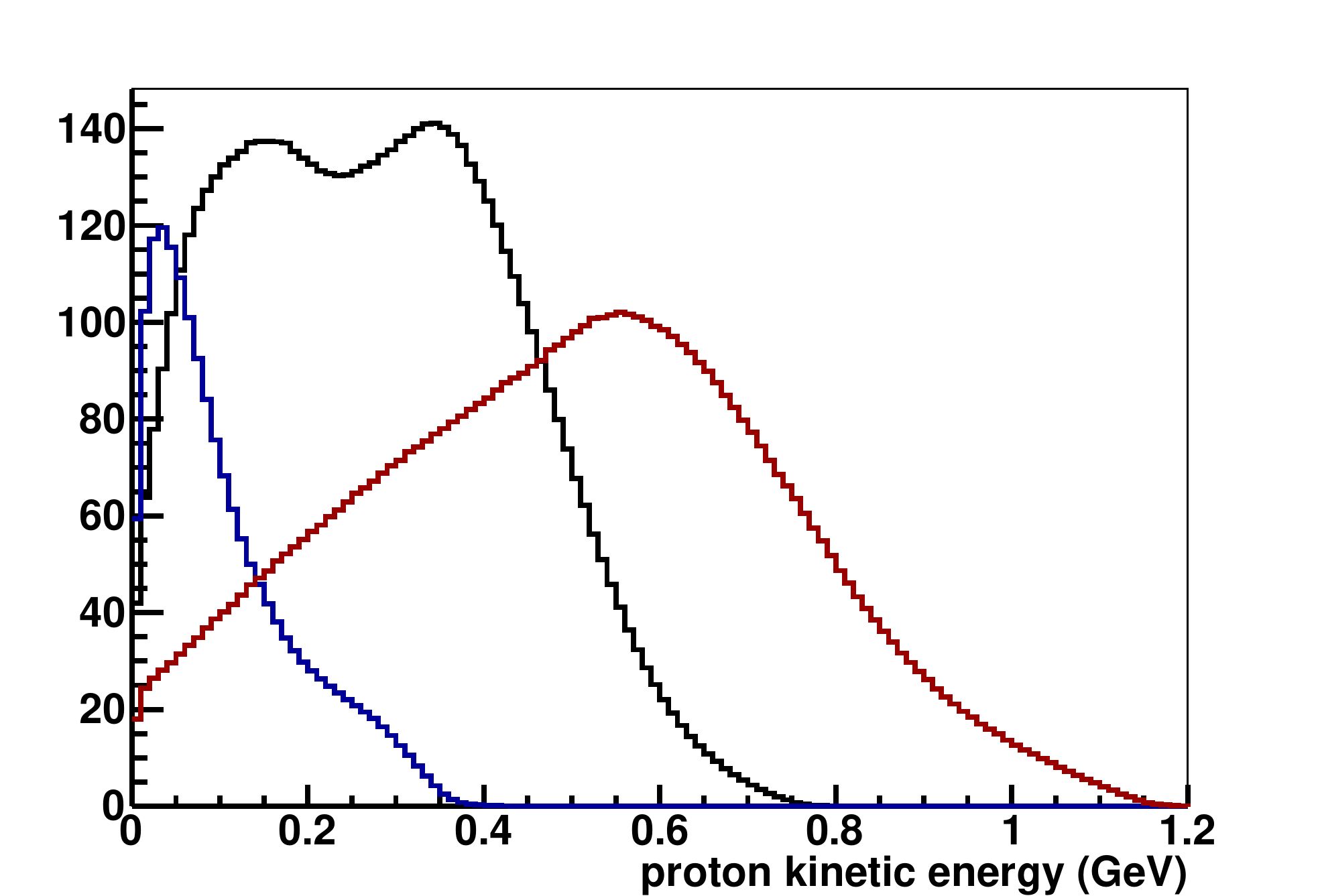}
\includegraphics[width=7.0cm]{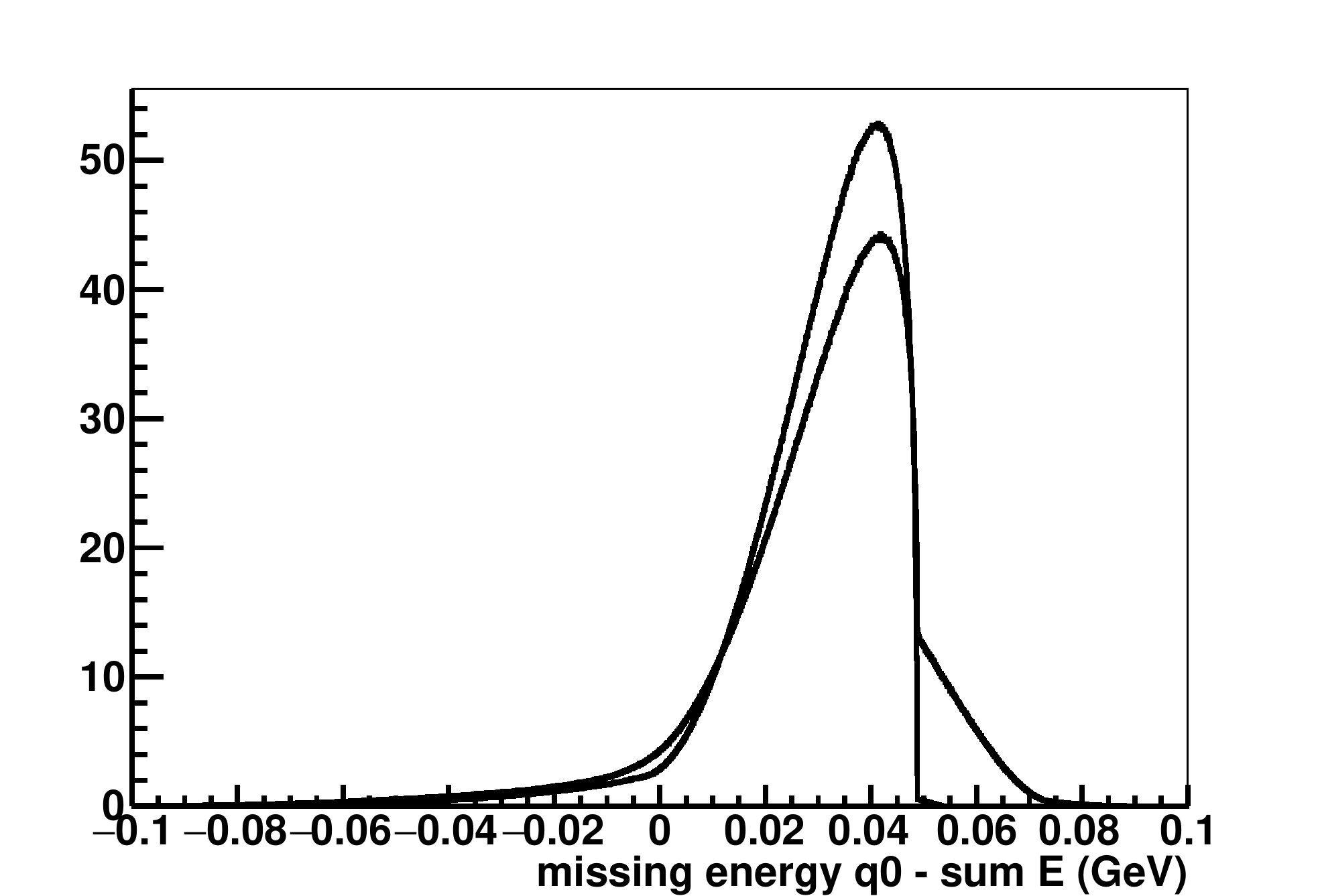}

\parbox{6in}{\caption{The available (i.e. sum the proton KE, but not the
    neutrons) energy on the left,
    for q3 $<$ 0.4 (blue), 0.4 $<$ q3 $<$ 0.8 (black), and 0.8 $<$ q3
    $<$ 1.2 GeV (red).  On the right is the missing energy, lost to
    unbinding nucleons in the interaction (with the sharp cutoff at
    0.05 GeV) and later during FSI (with a longer missing energy
    tail).
\label{fig:availableenergy}}}
\end{center}
\end{figure}

The plots in Fig.~\ref{fig:availableenergy} illustrate different aspects of the energy in the hadron
system, as might be viewed by a calorimetric or fine-grained tracking detector.  The left figure shows how much
energy will be easily seen in the detector as protons that can be
tracked and/or analyzed calorimeterically.  The black curve shows two
peaks which are a combination of the pn and pp final states as well as
the $\Delta$ and non-$\Delta$ contributions to the cross section.

The plot on the right in Fig.~\ref{fig:availableenergy} shows the
difference between the energy transfer and the sum of the energy of
the particles that exit the reaction.  The dominant feature is the
peak at 50 MeV, corresponding to the two units of nucleon removal
energy.  The curve with the sharp cutoff is before FSI, and broadened
one is after FSI.  Both curves have another tail on the negative side corresponding
to extra energy, which I think comes from striking nucleons with
unusually high momentum.

\section{More Technical Details}

\subsection{Hadron tensor interpolation}

The hadron tensors are read into Genie's BLI2DNonUnifGrid
class, whose Evaluate() method does a bilinear interpolation.  The
original IFIC Valencia model code also does a bilinear interpolation.  
The bilinear interpolation algorithm applied to the tabular hadronic tensor is
different than the one the original authors used;  it
interpolates beyond the first entries in the table, as if there is an
entry beyond that with a value of zero. This leads to slight differences in the resulting cross
section.  But the interpolation method is separate from the physics of
the process, so we have allowed that difference to remain.  Another
suggestion from the original authors is to use a 2D spline instead of
bilinear interpolation, though it seems unnecessary to achieve the 1\% accuracy.

\subsection{Hadron tensor resolution effects}

The hadronic tensor can be made with different step sizes. 
Historically, Nieves and Vicente's code produces it with 120 steps between 0
and 1.2 GeV in q0 and q3, which is used in this implementation also.
The resulting tensor requires two days
on a 2014-era 4-core machine using MPI.  Gran generated a tensor with
240 steps between 0 and 1.2 GeV, so four times the number of data
points, which requires about eight days.  These tests use the same
interpolation code, just different input tensors.

\begin{figure}[htbp]
\begin{center}
\includegraphics[width=7.0cm]{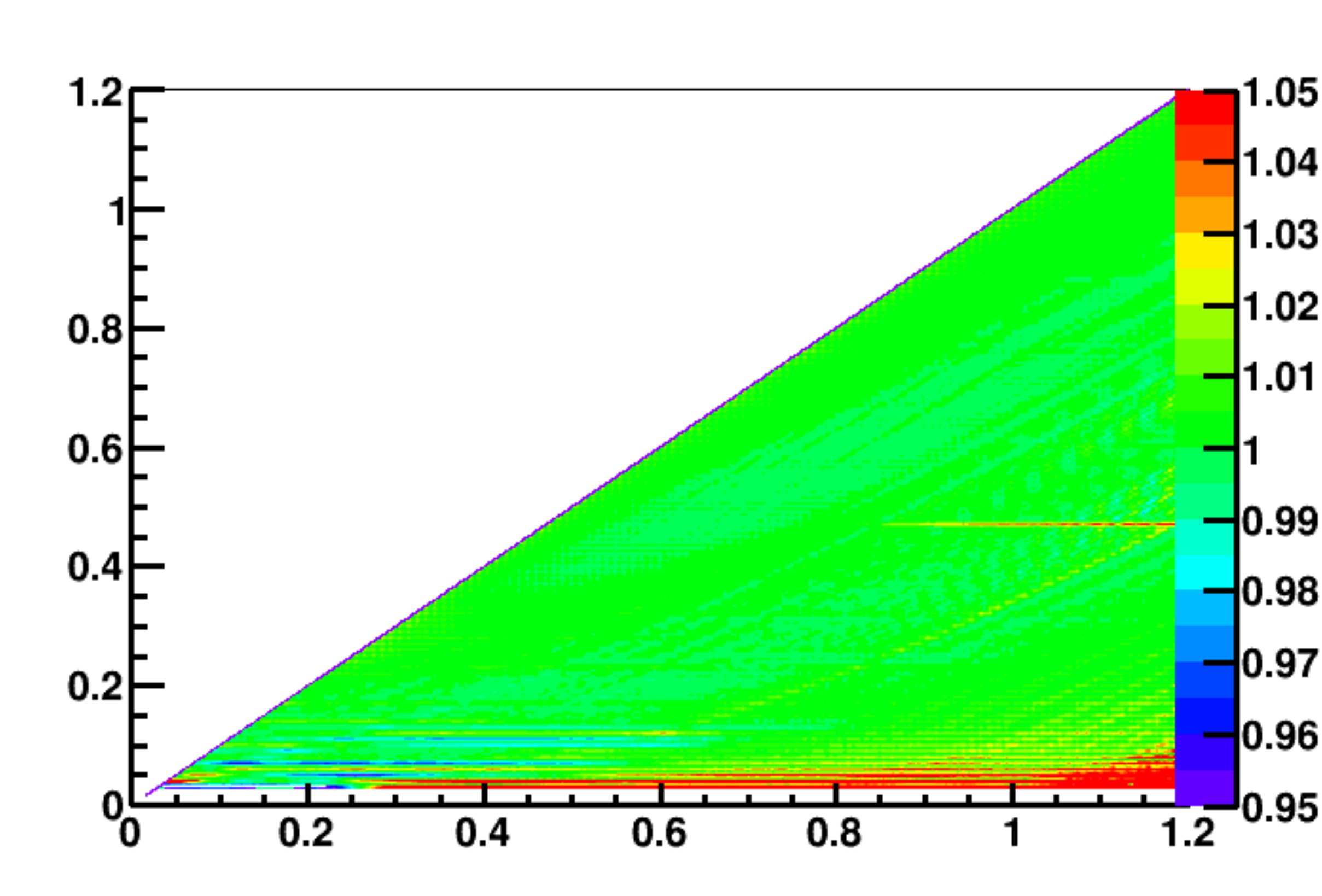}
\includegraphics[width=7.0cm]{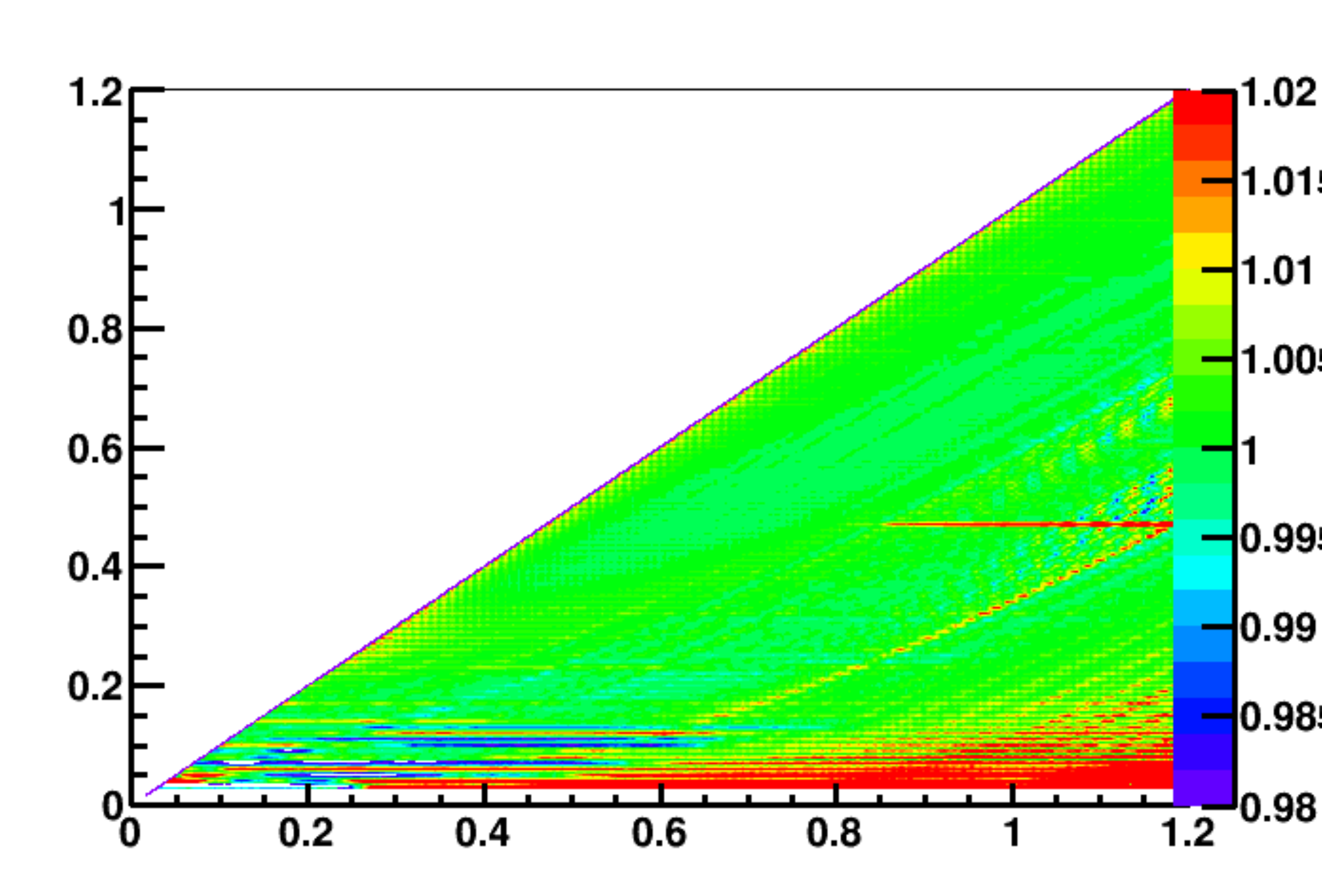}

\parbox{6in}{\caption{Effect of switching from a 240x240 HadTensor to
    a 120x120 HadTensor.  The top shows the color axis with a $\pm$5\%
    range, the bottom shows it with a $\pm$2\% range.
   The artifact at the top right is basically
     right at 5\%.  [Flatten these figures to reduce the filesize...]
\label{fig:hadtensorsteps}}}
\end{center}
\end{figure}

Differences appear at the 2\% to 5\% level, and none follow the main
lines of high cross section.
These tests use the same interpolation code, once the hadronic tensor is
read in.  It is not clear whether the few percent wiggles, or the horizontal
artifact just below $q_0$ of 0.5 GeV is a problem or failure in one or
the other calculations.  It might simply be riding the numerical
precision of the underlying cross section integrations.

The differences start to appear as you go lower and lower in neutrino energy,
producing a $\sigma(E)$ that is 1\% different or worse as you go
below 250 MeV neutrino energy.
Since the differences are small, it makes sense to use the smaller
files.  What is potentially a more interesting use of CPU power is to increase the steps
and precision in some
of the inner integrations and watch for effects at high $q_3$.

\subsection{Hadron tensor as text file}

The hadron tensor is provided as a plain text file containing 5 x 120 x 120
double-precision numbers.  For GENIE, we have reduced the file size by
truncating the {\small FORTRAN} output text precision to five significant figures and
representing zero with a single digit using a
python text processing script provided by Phil Rodrigues.  This step
reduces the file size by a factor of five.  When
tested, the resulting distributions are within 0.5\% everywhere which had
statistics good enough to tell, so this is as expected.  

Another choice would be to save the HadTensor as a binary file, either
of 8-byte (or even 4-byte) numbers, which might be rapidly read in, and be
similar or smaller in size.  Not sure this is relevant.

The text files could also be distributed as a tar.gz or tar.bz2, and
unpacked just in time.  For the seven nuclei here, the size is a little
over 3 MB.  These files are only read into memory once at the
beginning of generation, these additional steps might not offer any helpful optimization.

\subsection{Smooth turnoff of cross section at 1.2 GeV}

The implementation would benefit from a smooth turnoff of the cross section at
three momentum transfer of 1.2 GeV, otherwise some resulting
distributions will have a cutoff artifact.  Currently this is not
done.  Probably the way to do it
is to roll off the cross section linearly between 1.0 and 1.2 GeV.
Another way is to regenerate the hadron tensors out to 1.5 GeV, and
roll off the cross section smoothly between 1.2 and 1.5.   The GENIE
implementation can easily take alternate hadron tensors with larger or
smaller range.   Some experimenting with this is underway, to support
an estimate of acceptance and feeddown in analysis of data all the way
to $q_3$ = 1.2 GeV.
In any case, take care if your use of this model is
sensitive to an artificial step in the cross section.

The reason for the cutoff at 1.2 GeV is that cross section model itself
has a shortcoming stemming from its origins as a sub-GeV calculation.
The cross section does not
go to zero fast enough as $q_3$ $\rightarrow$ infinity, and the
integral might effectively or actually diverge at very high energy.
Also, the effect of higher resonances should become more important,
but are not incorporated into this model.
And at high $q_3$, the dominant process is neutrino-quark scattering
or deeply inelastic scattering (DIS), not neutrino-nucleon scattering,
and there are ideas about quark-hadron duality
that remind us not to double count such processes or nuclear effects.
Truncating the cross section at $q_3$ = 1.2 GeV preserves the structure at
low energy and momentum transfers without this divergence, and gives
access to the physics of primary interest.  When integrating higher
energy transfers, the DIS process will overwhelm what little cross
section is being left out except at high Bjorken $x > 1$.  

Actually its possible that an experiment
could look for events at very low energy transfer but very high momentum
transfer, above where this model cuts off.  Such events would
be at or below the QE line.  Even in this model, the band of cross
section that follows the QE line the closest happens to not follow a
line of constant W$\sim$1.0 GeV, but roughly crosses under the QE line.  At
about that point, the real cross section should be a combination of 2p2h
and 3p3h events, but also 1p1h events where the struck nucleon is part
of a SRC pair and has radically non-Fermi-gas momentum.  This part of
the cross section is not included with the hadron tensor.




\subsection{the max cross section in muon momentum and angle}

To optimize the accept-reject method, we have pre-computed the maximum
cross section as a function of neutrino energy.  Because we are
throwing for the cross section in the p$_\mu$ $\theta_\mu$ parameter space, the
maximum increases exponentially as a function of energy as the lepton
is boosted more and more forward for a given $q_0$, $q_3$.  The following
parameterization works for the neutrino-carbon interactions in the
IFIC Valencia
model.  It is multiplied by A/12 for all other nuclei.  Rather than
optimize and parametrize anti-neutrino and Delta special cases separately, we simply
use this largest parameterization in all situations.

{\footnotesize
\begin{verbatim}
double XSecMaxPar1 = 2.2504;
double XSecMaxPar2 = 9.41158;
int NuclearAfactorXSecMax = 1.0;
  if (TgtPDG != kPdgTgtC12) {
    if (TgtPDG > kPdgTgtFreeN && TgtPDG) {
      NuclearA = pdg::IonPdgCodeToA(TgtPDG);
      // The QE-like portion scales as A, but the Delta portion increases faster, not simple.
      // so this gives additional safety factor.  Remember, we need a safe max, not precise max.
      if (NuclearA < 12) NuclearAfactorXSecMax *= NuclearA / 12.0;
      else NuclearAfactorXSecMax *= TMath::Power(NuclearA/12.0, 1.4);
    } 
  }
double XSecMax = 1.35*TMath::Power(10.0, XSecMaxPar1 * TMath::Log10(Enu) - XSecMaxPar2);
if(NuclearA > 12)XSecMax *=  NuclearAfactorXSecMax;  // Scale it by A, precomputed above.
\end{verbatim}

}

The maximum cross section in $q_0$, $q_3$ is relatively stable.  In
principle, the code could instead throw for this and then compute the
lepton kinematics.  This would permit a single constant of about
$7\times10^{-38}$ cm$^2$/GeV$^2$ for carbon, and some scaling by A to work for
the IFIC Valencia model.

This procedure is ``safe'' in the sense that an assert( ) call will
require the cross section obtained from the hadronic tensor be less
than the specified maximum cross section, or generation will spout an
error and crash.  This has of course been tested for all the default
tensors, but might happen if you replace one with your own.  Its also
not especially optimized, generation of Pb208 events takes longer than
it needs to, for example.


\subsection{throwing for the nucleon pair}

Two nucleons are pulled from whichever GENIE nuclear model the user
configured.   As of 2015, almost everybody using a generator uses the global Fermi gas
(with or without the Bodek-Richie tail),
though the IFIC Valencia model uses a local Fermi gas in its calculation, and some are
enthusiastic to use a spectral function.  Since the hadron system is
already not true to the underlying calculation, our implementation
takes the nucleon momenta from the prevailing nuclear model.  This
includes the removal energy attached to that model, which might be a
constant for each nucleus or might change with each nucleon drawn.

\begin{figure}[htbp]
\begin{center}
\includegraphics[width=5.0cm]{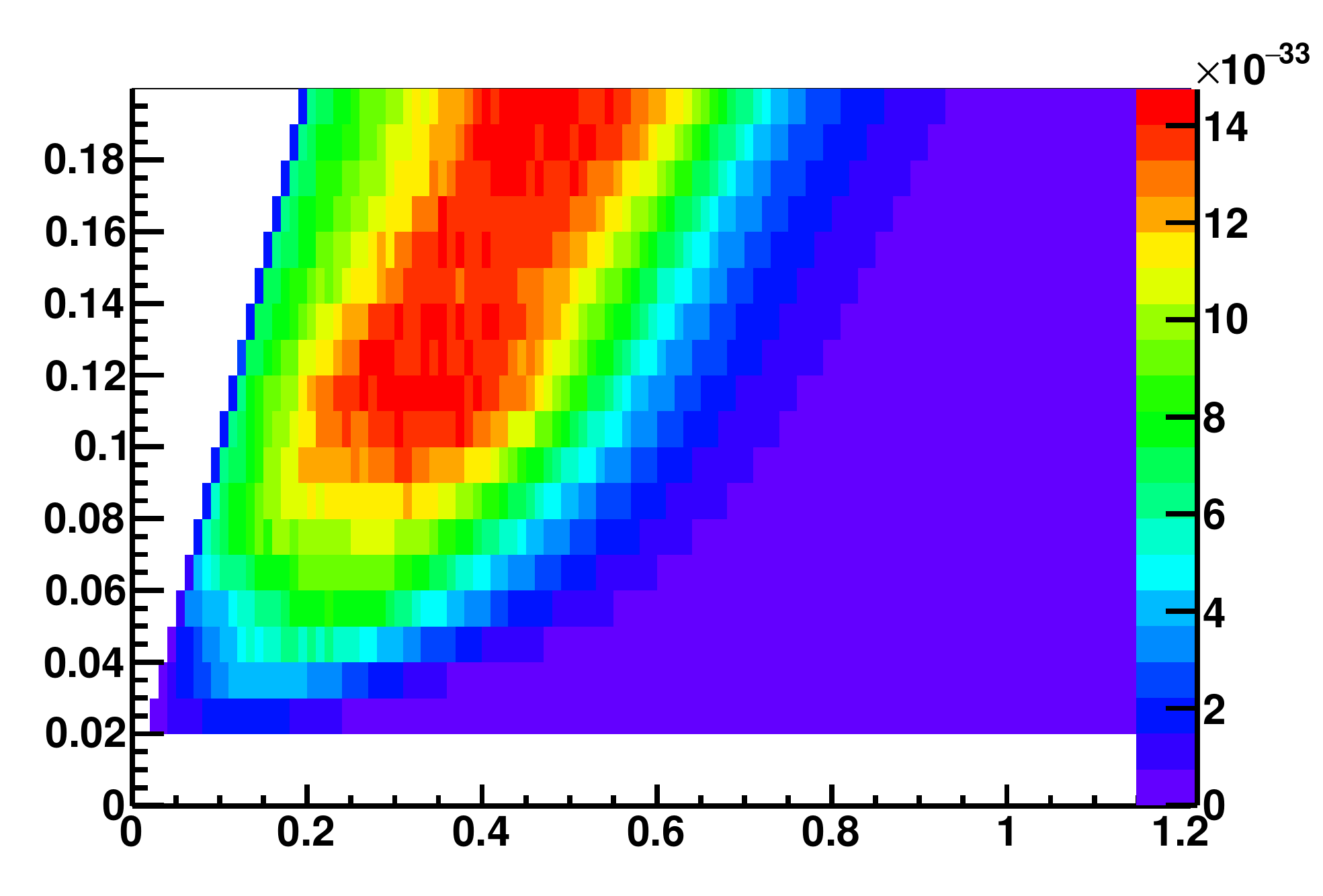}
\includegraphics[width=5.0cm]{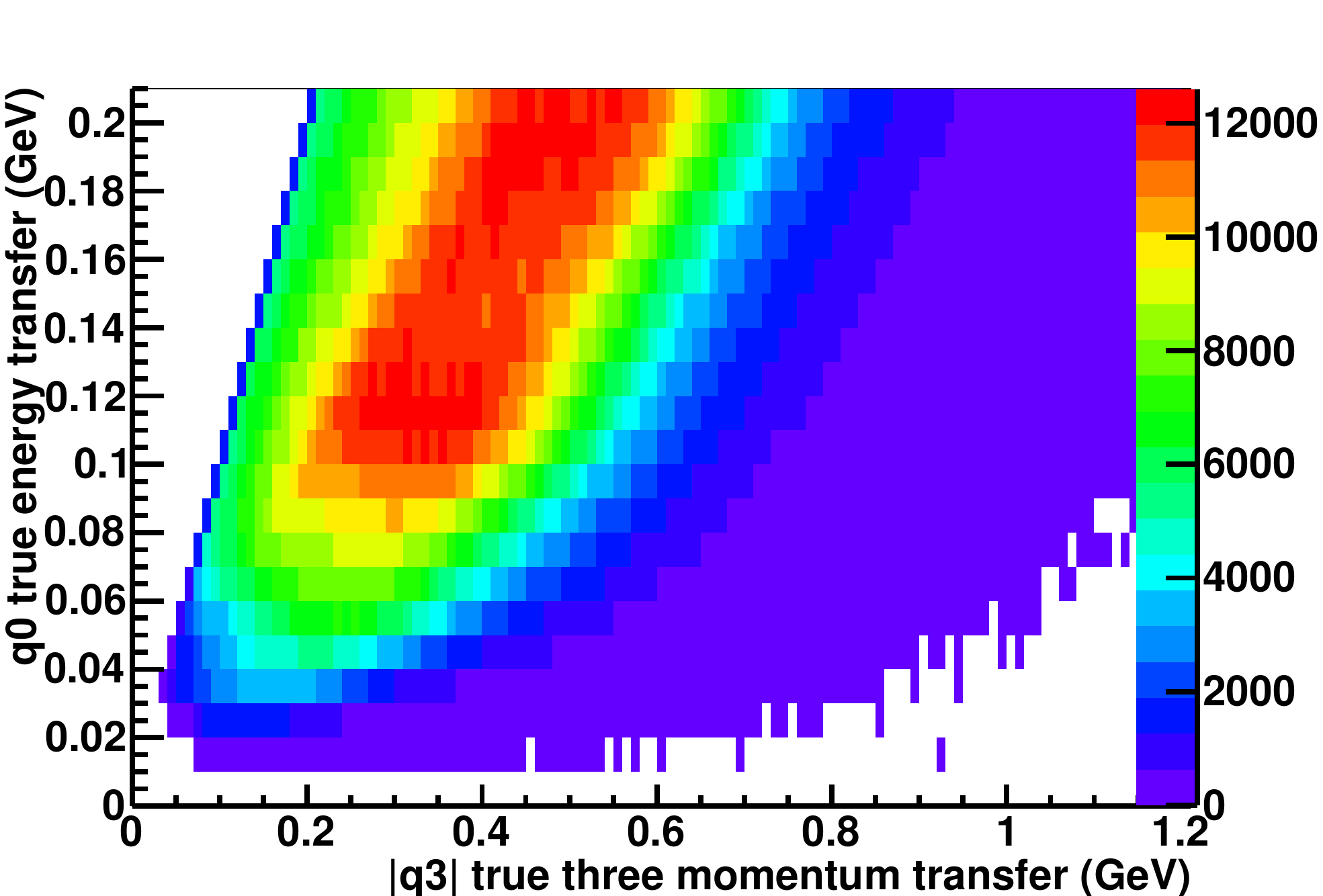}
\includegraphics[width=5.0cm]{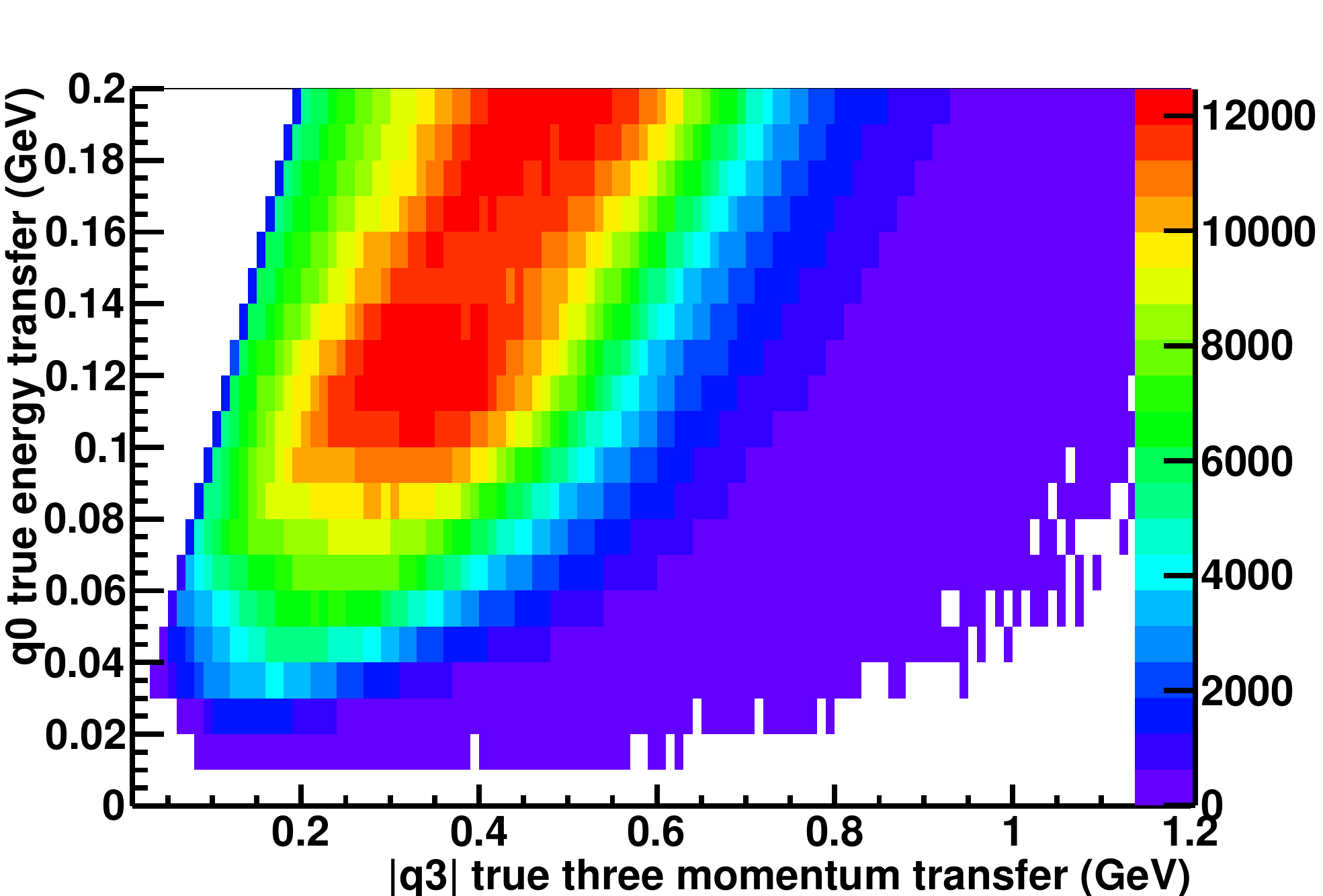}

\parbox{6in}{\caption{Comparison of the relative event rate between
    the IFIC Valencia calculation (left) and two versions of GENIE one with the
    default 1000 iteration limit (right) and one with 100000 iteration
    limit for throwing the hadron kinematics.  As you go from right to
    left, the lower portion of the event rate is systematically shaved
    off more and more and more These are
    neutrino-carbon 3 GeV.  The statistics of the two GENIE samples
    are the same, the statistics of the Nieves calculation is much
    higher, so the lower right corner of the plot is completely filled
    in statistically.
\label{fig:mecpoint}}}
\end{center}
\end{figure}

The place where we are throwing for a nucleon pair iterates if the
resulting pair (after momentum and energy transfer and binding energy)
is off shell.  Especially at very low energy transfer corner of the
triangle of kinematics,
this loop will sometimes terminate without finding acceptable
kinematics.  Termination kicks an error up the chain and causing GENIE to rethrow the
entire interaction.  Too much of this causes a bias in the resulting event rate,
compared to the IFIC Valencia cross section.  This is illustrated in
Fig.~\ref{fig:mecpoint} by a sequence of three 2D event-rate plots.

As a temporary mitigation, we have
multiplied the global iteration max kRjMaxIterations ( default = 1000
) by 1000 for this specific loop, an earlier test with 100 times
default is illustrated in Fig.~\ref{fig:mecpoint}.  The left plot has a high
statistics event rate distribution from the IFIC Valencia modle calculation directly.
The other two are 50M stats samples from GENIE which allow 100,000
(center) and 1,000 (right) rejections.  The sculpting is visible in the lower
left corner of each plot, and in the lower right corner of the
right-most two (equal-stats) plots.

Ordinarily the solution is to impose a kinematic maximum as a
function of the energy transfer, and throw for nucleons within that
maximum, resulting in acceptably small iterations even at this
kinematic limit.  In this case, we are using a method from
the nuclear model class, so we don't see an easy way to impose a limit
from within the MECTensor model.  

Caution:  switching to even more wacky nuclear models (local fermi gas
or spectral function) might exacerbate the problem caused by
terminating this loop.


Of course, we are not changing the total cross section at all, because
that is set by the spline.  What we are doing is enhancing the rest of
the kinematic region's double differential cross section by an amount
that is less than 0.01\% of the total, but getting this corner wrong
by 1\%.  This seems acceptable.

One other artifact is visible in Fig.~\ref{fig:mecpoint}.  The GENIE
based plots have event rate below $q_0$ = 0.02 GeV, but the IFIC Valencia model
calculation does not.  This is coming not from physics, but from the
2D interpolation method.  The model authors coded it in a way that gives a limit at
the first bin of the table, which corresponds to $q_0$ = 0.021 GeV for
carbon.  The GENIE interpolator assumes the table goes to zero just
outside the bounds of the table, and is willing to interpolate into
this.  In principle this affects both the spline and the differential
cross sections by about 1\%.




\subsection{Removal energy when generating the hadron system}

The IFIC Valencia model lepton tensor calculation considers particular choices about the CC
process Q-value for
protons or neutrons and the three isoscalar nuclei $^{12}$C, $^{16}$O, $^{40}$Ca in
Table I of the QE+RPA paper \cite{Nieves:2004wx}, or from any table of
nuclei.   With the incorporation of this model, we updated GENIE's
model for nuclear masses to the latest complete compilation
\cite{Wang:1674-1137-36-12-003}.  For some common remnant nuclei like
C11,  this was a change of 10 MeV and has a Q-value-like effect on the energy of
the outgoing products, especially the lepton for QE events.

Beyond that, the Valencia code integrates over the specifics of initial
and final hadron states, so that aspect of the model is not available
to us.  In this case, the GENIE version of those specifics are substituted.

The initial state hadrons are not drawn from the
same local fermi gas used in the Valencia model.  We follow the user's choices for GENIE's
nuclear environment.  The method here almost matches how GENIE treats
removal energy in the CCQE case, and is slightly different than what
NuWro does.  The nucleon cluster is formed from two nucleons drawn randomly from
the prevailing nuclear model. 

Most nuclear models in GENIE also
assign a removal energy to nucleons.  The default Fermi-gas model
considers the removal energy a constant 25 MeV for carbon, regardless
of the momentum assigned to the nucleon.  Alternate models such as a
spectral function draw a removal energy explicitly from a probability
distribution function when choosing a
nucleon.   In contrast, a local Fermi-gas would correlate the removal energy
with the local nuclear density.

These two quantities (one for each removed nucleon)
of removal energy are subtracted from the energy
transfer and added to the remnant (excited) nucleus.  In the versions
of GENIE up to now (2.12.6 and earlier), this energy is unavailable and
is not transferred to the Geant4 system for most or all nuclei.
In the case of NuWro \cite{Sobczyk:2012ms},
 this same thing is done, but after the di-nucleon cluster is split, then an additional
8 MeV of binding energy is removed from each nucleon ``to put it on
shell''.  

GENIE technical point.   There is a second mechanism for accounting
for where removal energy within GENIE ends up.   It can attached to a
neutron or proton using the SetRemovalEnergy method.  When the event
is complete, all the final particles are queried to determine if there
is a stored removal energy, and they are placed in the event record as
a final state particle called ``NucBindE'' with the removal energy
(e.g. 25 MeV) appropriate for that nucleus.  This method is only used
for protons and neutrons, and are visible especially in quasi-elastic
events.

GENIE physics point.   The above two methods lead to the same result:
there are one or more particles in the final state that will not usually lead
to energy deposit or deexcitation, but should and may contain 25 to 100 MeV or
even more energy.   At this point in GENIE, it doesn't matter whether
its in the form of a HadBlob or NucBindE, it will be ignored when
passed to Geant4.  There are plans within GENIE to improve this.
The MINERvA data on the calorimetric hadron system
\cite{Rodrigues:2015hik} has some sensitivity to the unsimulated portion of this
energy that appears promptly in the form of additional ejected
nucleons or deexcitation photons if they happen within 35 ns or 150 ns, 
depending on the specifics of the analysis.   (MINERvA would also be sensitivite to
radiative processes producing a photon and modifying the lepton
kinematics).


\subsection{Decay the nucleons}

The implementation calls ROOT's TGenPhaseSpace class, which was based
on the GENBOD W515 n-body event generator.  The ROOT class further
mentions it uses the Raubold and Lynch method, and is also documented in
F. James, Monte Carlo Phase Space, CERN 68-15 (1968).

Most information about the separate initial states of each nucleon is lost,
only the combined three-momentum and energy, and the isospin state, is
passed to the phase space decayer.  This is obviously a simplification
that ignores details at are buried in inner integrations of the IFIC Valencia model calculations.
We have chosen to live with for this early implementation.

This class will make a two-particle system back-to-back and isotropic in the center of
momentum frame, then boost it back into the ``lab frame'' of the
particle that was passed to the class.  It returns a weight, which
must be queried with a random number followed by rethrow if the random
number test fails.  This weight presumably doesn't matter (?) for
isotropic two-body decays, but gets the distribution of three-body
decays right.  

Using this fancy isotropic decayer is unnecessary if all we ever
cared to do was two nucleons.  In principle, the code is prepped to extend this model
to include some 3p3h states, and still use this phase space decayer, which does
appropriately more interesting things in this case.

\section{What to do about the Delta component}
\label{sec:delta}

The calculation is NOT of an inclusive 2p2h process, it is of the
QE-like (no pion) component of the 2p2h process.  Even then, it
excludes the QE-like 1p1h with SRC diagrams, which Nieves instead incorporates
in his RPA model.  

The hadron tensor explicitly
includes a portion of the cross section with Delta kinematics, with a
Delta produced at the diagram level which ``decays'' back to a nucleon
during the eponymous meson exchange, or maybe the
$\Delta$ is ``absorbed'' at the diagram level.  This
happens before any $\Delta$, pion, or nucleon is passed to the Final
State Interaction model.  In this model, such delta events still have
an ``experimenter's W'' invariant mass of approximately 1.232 GeV.

The total calculation of the amplitudes includes terms with and
without the Delta, but not with higher order resonances.  There are
significant interference terms that do not change the cross section
compared to ignoring the Delta amplitudes, but do significantly change where in the
kinematics the cross section sits, and where the pn initial states are the
highest.

About 16\% of the deltas produced by the GENIE default model already lose their pion
in the FSI model to some absorption/knockout process, leading to two
or more nucleons in the final state.  A similar
number of events are predicted by this 2p2h model to
have delta kinematics but with two nucleons and no pion, before FSI.
A core prediction of the model is that there will be more nucleon-only
events in a QE+delta kinematic sample than a prediction without this
process.
Other generators, such as NEUT, have an available
pionless delta decay mode.

The effect of the interference is clear for the QE-like sample, but it
is not so clear for the inclusive sample.  Do the pionless delta
decay/absorption events take the place of 1p1h (and/or 2p2h) delta events with a
pion, or are they in addition to those, or somewhere in between.

Given the nature of the prediction, it makes sense to ask this model
to generate all the 2p2h pionless delta component, and pass the result
to FSI.  Simultaneously, the user should generate 1p1h deltas and pass
their products to the FSI model, some of which will lose their pion.
Then the user (concerned about model-dependent systematic effects)
might consider the following two extremes.  One is to keep the 1p1h
delta strength as it is (informed by deuterium data and other
evidence), the other is to reduce the 1p1h delta
strength by the amount of 2p2h pionless delta produced.  An inclusive
calculation would presumably yield something in between, and/or have
something specific to say about the isospin states involved.

The facility to do the above can be accomplished because the model
presented here includes the Delta-only hadronic tensor, and can
calculate the resulting cross section on its own.
With that, the presence of the Delta in the generated can be
tagged statistically for each interaction with the internal computation of two
more cross sections and additional random throws.  On an
event-by-event basis, this is saved as the Delta resonance type
(xcls.Resonance() == 0) ;
This allows the user the option to generate the events from the total Nieves
model,
but then weight the Delta-induced 
component to zero in favor of their other favorite pionless-Delta
model, like the one that has been available in NEUT or a weighting up
of the FSI version.   Note, users of this version of GENIE who already
(from past versions)
use the resonance ID method should also query the interaction type for
resonance or 2p2h, if that distinction is important for analysis.

\section{What to do about other nuclei}
\label{sec:othernuclei}

The implementation in GENIE from 2.12.6 onward generates interactions
all nuclei with A $\geq$ 9, gives a rough guess
for helium-4 (needed today only by MINERvA).   It does this by creating a
set of hadron tensors for a few specific nuclei that span the periodic table
and scaling to any nearby nuclei the user requests.

\begin{table}
\begin{center}
\begin{tabular}{c|c|l}
He4 & $\sim$ from carbon 12 & MINERvA special \\
other A $<$ 9 & not available & \\
9 $\leq$ A $<$ 15 & carbon 12 & will give N14\\
15 $\leq$ A $<$ 22 & oxygen 16 & \\
22 $\leq$ A $<$ 33 & silicon 28 & will give Al27\\
33 $\leq$ A $<$ 50 & calcium 40 & will give Ar40, Ti48 \\
50 $\leq$ A $<$ 90 & Ni56 = pseudoFe & will give Fe \\
90 $\leq$ A $<$ 160 & Ba112 = pseudoCd & \\
160 $\leq$ A  & Rf208 = pseudoPb & 
\end{tabular}\\
\parbox{6in}{\caption{Hadron tensor used for the range of nuclei in
    GENIE.   The PseudoA means the code used an isoscalar nuclei with
    the density parameters of Fe, Cd, and Pb.  
\label{fig:nuclei}}}
\end{center}
\end{table}

The IFIC Valencia model and the hadron tensors are based on isoscalar nuclei.
However, the version of the model that uses hadron tensors
approximately scales with A, and combinatorics can be used make a
prediction for nuclei that are not isoscalar.


The strategy is to start with isoscalar nucleus with A near
each of the nuclei we expect users will want.  Then a good
approximation is to scale the pn (and the not-pn) initial fraction to
account for the different number of protons and neutrons in a nucleus
of the same A.   That same scaling will naturally account for a
slightly different A, as long as it is not too different.

For example, to get $^{40}_{18}$Ar from $^{40}_{20}$Ca, which have the
same A,
we should first take the
probability that we had an initial pn pair in Argon (22x18) divide by
those in Calcium (20x20), and use that as the scaling factor 0.99.  In
fact, this prescription would cause the cross section to scale as
$A^2$, which is not desired, so taking the
square root returns a factor even closer to 1.0.
Likewise, scale the initial nn pairs by Sqrt( (22x21/2) / (20x19/2) ) = 1.10 
and apply those scales to the appropriate components and reform the
total cross section.  Something similar for initial pp pairs in the
anti-neutrino case, (18x17/2)/(20x19/2) then the square root yielding
0.897. 
This strategy extends to one other vital nucleus, where we could approximate
$^{56}$Fe scaling from $^{56}$Ni.

Higher on the periodic table, there are no naturally occuring
isoscalar nuclei on which to base the hadron tensor.   The calculation
was asked to produce tensors for barium 112 and Rutherfordium 208, but
was told to use the nuclear density parameters for cadmium and lead
respectively.   Even the example above, using iron is more likely than
nickel, so the hadron tensor for isoscalar A=56 was generated with the
iron density parameters, though they are not very different.
Then the scaling away from isoscalar proceeds as
before.  That is quite a big scaling from Rf208 to Pb208, and it is
difficult to evaluate the accuracy of the result, except it is certainly
a better approximation than having no QE-like 2p2h component at all.

The author's code
uses nuclear density parameters to approximate these nuclei, which is
not tested and may not be appropriate for helium.
Neverthelsess, He4 is added for the 2017 release of GENIE 2.12.6, 
to satisfy the immediate need for some generated events from helium 
for use with data from the MINERvA experiment.
The code will generate events for helium using the above prescription,
and the carbon hadron tensor as input, plus the actual Q-value for
helium.  This will produce 1/3 the events on pn pairs, and the
combinatorics will yield 1/4 the events on pp and nn pairs. 
It is not suitable to consider these events as
predictions of the model, and any study doing so should state so. 
But the resulting events could be used to
evaluate acceptance and resolution effects and study potential
sensitivity to future models.  
GENIE will not produce events on D, T, He3, nor Li based on this model.

In the 2013 paper \cite{Gran:2013kda}, it was reported that the Delta
component did not scale exactly with A.   That paper used two
variations of the calculation, and the latter variation that separates
out the
hadron tensor calculation does not have that strong, non-A scaling behavior.   
Therefore, the GENIE
implementation also scales very nearly with A.

This prescription leads to three kinds of differences between any two
nuclei.   The scaling of pn, pp, nn is clear from the discussion
above.
Another comes from the applicaiton of Q-value.  The third
comes from the choice of nuclear density parameters, if different
nuclei are widely separated on the periodic table.  

\begin{figure}[htbp]
\begin{center}
\includegraphics[width=5.0cm]{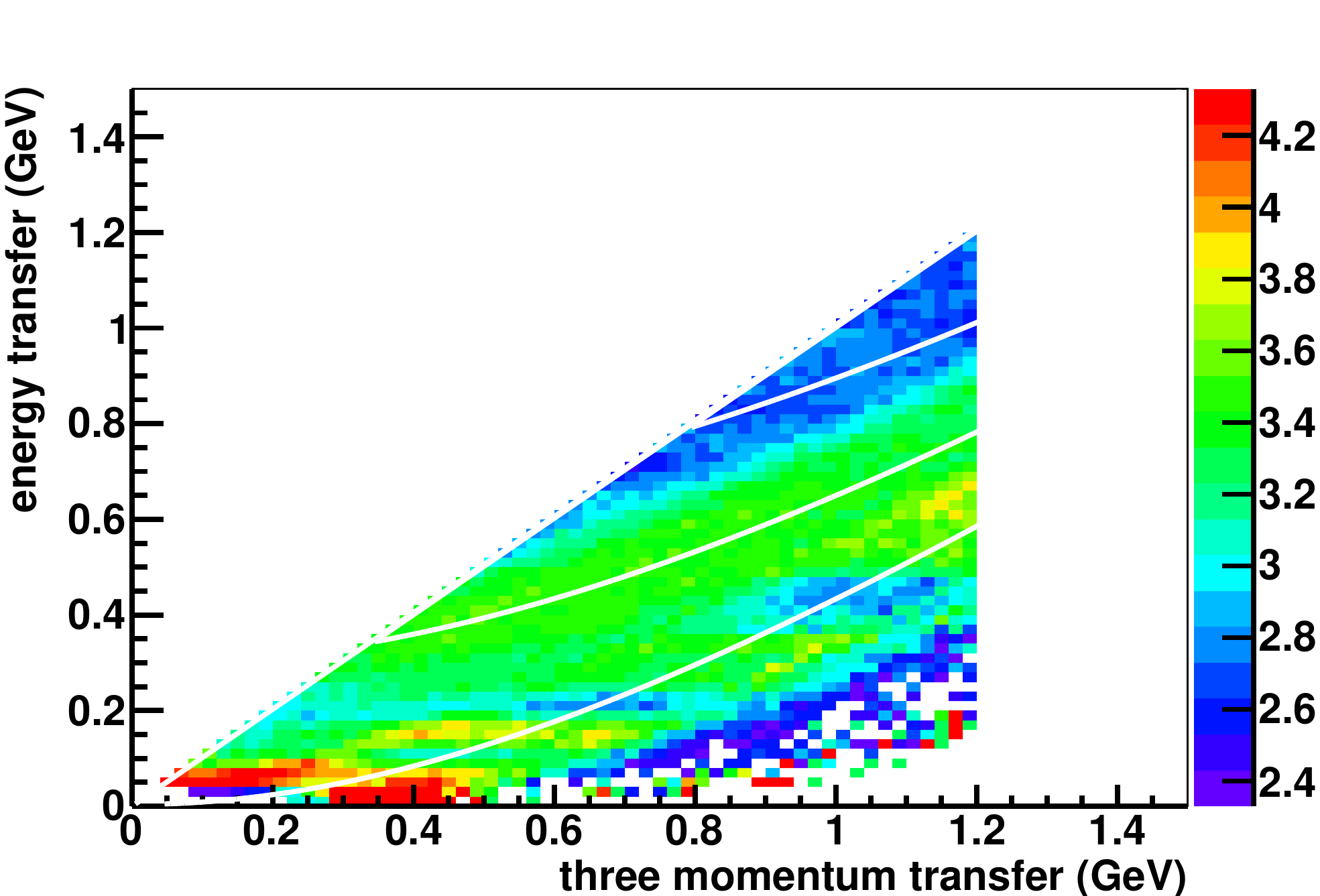}
\includegraphics[width=5.0cm]{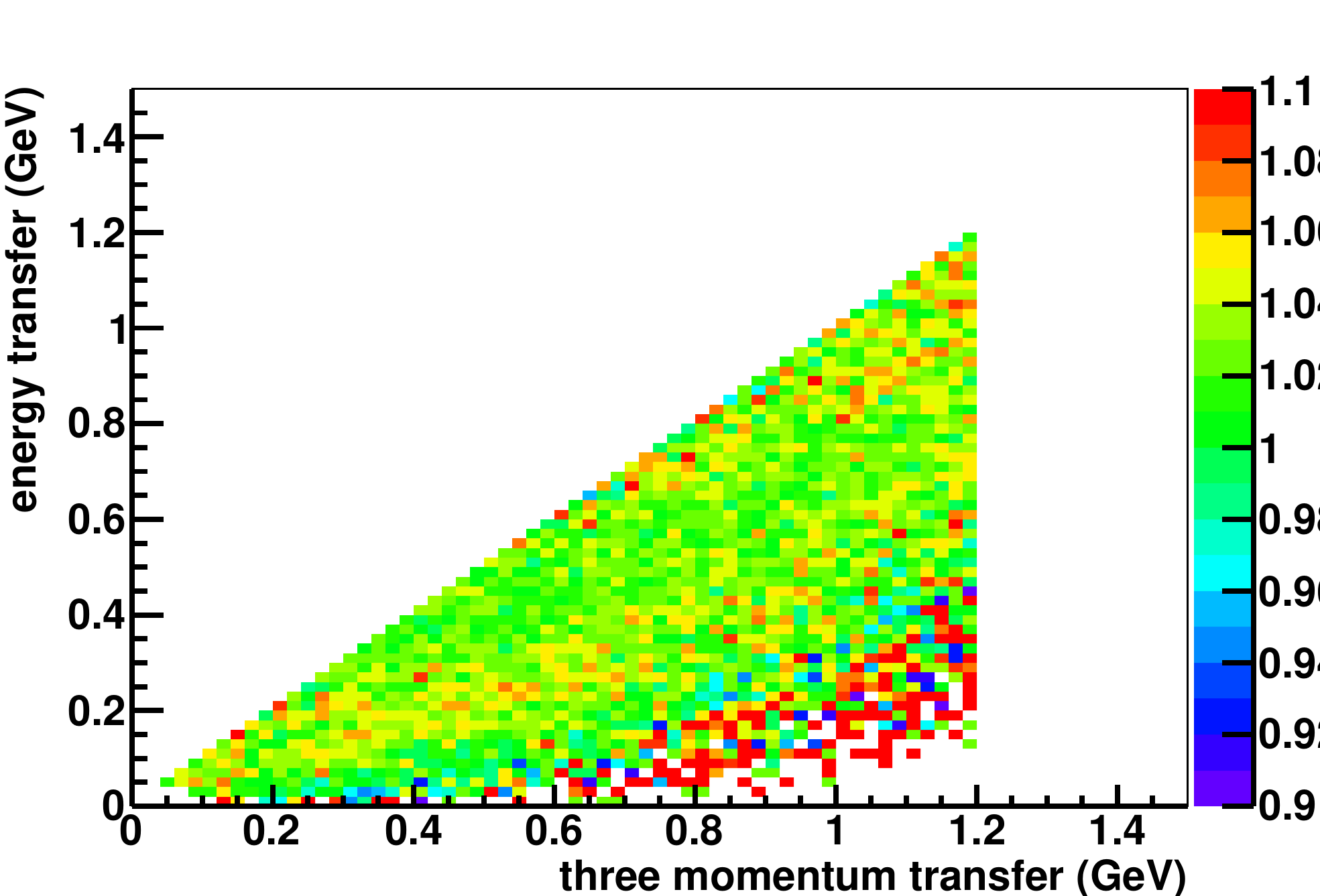}
\includegraphics[width=5.0cm]{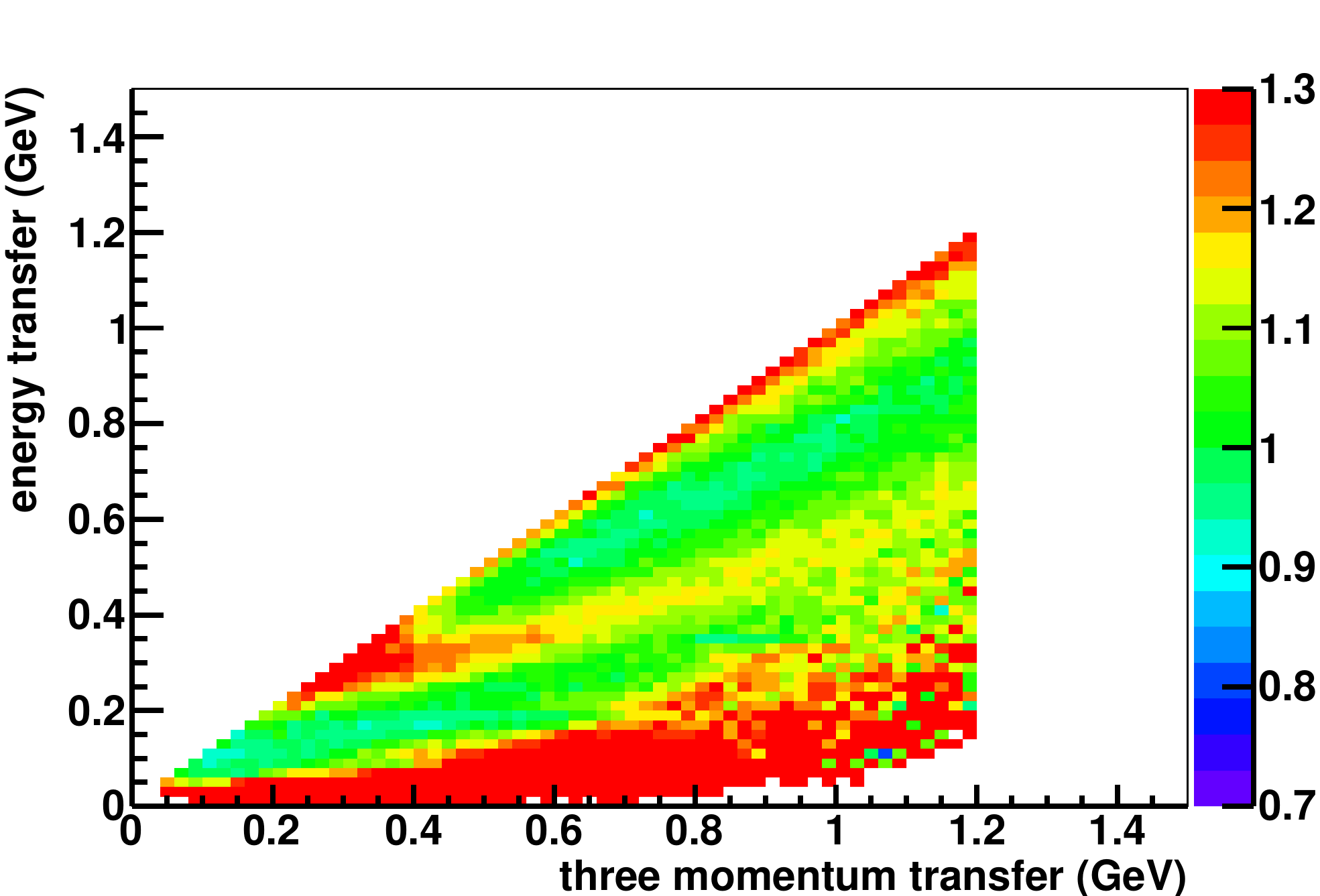}

\parbox{6in}{\caption{Comparison of the three components that come
    into play when comparing two nuclei.  The left shows the ratio of
    Ca40 over C12, forcing Ca40 to have the C12 Q-value, so only the
    nuclear density parameters and total A are different.   The middle
    shows the scaling away from isoscalar nuclei Ar40/Ca40, giving the numerator
    the same Q-value as the Ca40.   The slight statistically
    significant differences enhance the !pn component.   The right
    plot isolates the Q-value effect, computing Ar40/Ar40 with the
    denominator having the Q-value for Ca40 but no other differences.
\label{fig:Adependence}}}
\end{center}
\end{figure}

A comparison is made of the calculation using the nuclear density parameters
for Ca40 and C12, but both are isoscalar and the comparison
artifically sets the same Q value.   This comparison is in the left
plot of Fig.~\ref{fig:Adependence}.
This yields 30\% changes in the regions where
there was little event rate, and some 10\% artifacts in the QE region
that follow lines of the internal numerical integrations.  There
could be 5\% effects beyond A scaling in the pp and nn initial state reactions.

The most extreme changes in Q-value cause noticeable changes in the
cross section, shown in the far right plot of Fig.~\ref{fig:Adependence}. 
Despite being very nearby, stable nuclei, neutrino
reactions on Ar40 has Q = 1 MeV while Ca40 is 14 MeV.   This causes
the entire cross section distribution to shift down or up by that 13
MeV.  With the peaked distribution, this causes 30\% changes in
regions where there was little event rate to begin with, and -10\%
changes at the peaks.   This modification of the kinematic space
can also cause nearly 10\% changes in the integrated cross section.

The middle plot shows the effect of the pn and nn scaling to a nucleus
with relatively more neutrons in the initial state.   The statistics
of the sample, the smallness of the effect (Ar40 and Ca40 are near to
each other), and that its diluted from combining pn and nn at the same
time, mean its hard to see.  The trend indeed follows from the
upper right plot of Fig.~\ref{fig:diffxs}.



\section{Model comparisons, tuning, uncertainties}

There are many models on the market.   This section has been updated
to include the most prominent as of early 2017.  Apples to apples
comparisons between these models are rare, comparisons to the same
data sets are more common. 

GENIE has a W-parameterized model built by Steve Dytman
\cite{Katori:2013eoa} inspired by the Quasi-Free
Scattering model implemented in {\small FORTRAN}
of Lightbody and O'Connell \cite{LIghtbody:1988} in the 1980s.  This model was available in the early
days of JLab, and was used in proposals to
pursue 2N signatures in electron scattering data and other inclusive
reaction studies.
Like our need, this code included QE
scattering, two-nucleon knockout in the dip region, the Delta and the
next two resonances, and a DIS component.   The original version used
a Cauchy/Lorentz/Breit-Wigner distribution as would be normal for the $\Delta$
resonance.  In the current Dytman
implementation for GENIE, the MEC component is simplified a Gaussian
distribution in W centered on 2.1 GeV with a width of 0.5 GeV and
$Q^2$ peaked near zero.  This corresponds to 225 MeV of kinetic energy
available in the center of momentum frame once the nucleons separate,
and should place these events  
somewhere between the QE and the Delta, in the dip region of the usual
kinematic spaces.  Since GENIE version 2.10.6 it takes the fraction of
initial-state di-nucleon clusters to be 80\% pn, for both neutrino and
anti-neutrino.  Some of these parameters can be
configured to other values in the options file.    Also since this version, the cross
section no longer rolls off at 5 GeV by design.  This was a way to balance the
potential disagreement between the MiniBooNE and NOMAD results, but
that explanation was ruled out by MINERvA data \cite{Rodrigues:2015hik}.
This previous energy dependent behavior was similar to the proposal 
of the Transverse Enhancement Model inspired by inclusive electron
scattering data.

The model of Martini and Ericson \cite{Martini:2009uj} is the
successor to earlier work of Marteau \cite{Marteau:1999kt}.  Marteau's
work itself was based on prior work by Ericson an Delorme in the
1980s.  This overall effort is sometimes referred to as the ``Lyon
Model'', though here we'll refer to its main proponent and call it
Martini's model.  The diagrams they compute are
similar to the IFIC Valencia model, but as of 2014 were missing a few
diagrams that Nieves includes.   Some versions of this model are
non-relativistic, while other version contain progress toward a more
completely relativistic calculation.  Also, the overall cross section
predicted by Martini's model is about a factor of two larger than
the IFIC Valencia model, though both describe electron scattering data similarly.
Maybe it includes the SRC component that Nieves leaves to his RPA model.
Detailed comparisons of the two models at the range of energies of
interest are few and limited.  

A comparison of the Martini and
Nieves models and a subset of MiniBooNE data is in the
ICHEP2014 conference proceedings \cite{Nieves:2014lpa}, illustrating
the normalization difference, and suggesting the Martini model
produces events closer to the central QE kinematics (further from the
center of the dip) than the IFIC model does.  Some of the technical
differences between these two approaches are highlighted in the
introduction to \cite{Simo:2014wka}.

Additional progress has been made in collaboration with Martini and
Jachowicz and their students in Gent, especially Van Cuyck
\cite{VanCuyck:2016fab,VanCuyck:2017wfn} to accompany their work on
the quasielastic process.  Broad comparisons to other models are not
yet available.
The precursor to Martini's Model, by Marteau \cite{Marteau:1999kt}, is implemented in the
NuWro generator, and numerous predictions and comparisons to what is
presented in this note might be readily obtained if we ask the NuWro
experts.  There are a number
of NuWro talks on this topic, including comparisons with the IFIC
Valencia model.

A campaign has been launched by proponents of the superscaling (SuSA)
approach, including work by Megias, plus also Nieves and Vicente's former
student, Ruiz Simo.  This effort builds off the work by De Pace, et al. \cite{DePace:2003xu} 
which was a microscopic calculation of 2p2h effects in electron
scattering.  This previous effort augmented the authors work on the
SuperScaling (SuSA) description of the QE and $\Delta$ processes, so
they could describe inclusive electron scattering including a
component in the ``dip region'' that probably does not obey
superscaling.
Ruiz Simo is investigating the full computation of the 2p2h process,
adding the axial terms, and doing so without
the simplifications that yield a hadronic tensor with only four or two
dimensional integrations and 8 cpucore.days worth of computation.
Progress thus far has been described in
\cite{Simo:2014wka,Simo:2014esa,Megias:2014qva,Simo:2016ikv,RuizSimo:2016ikw,Megias:2016fjk,RuizSimo:2017onb,Amaro:2017eah}, which cover a mix of physics and
computational challenges, with careful but limited comparisons to simpler or
asymptotic approaches. 

The transverse enhance model \cite{Bodek:2011ps}
extracts an enhancement from
(e,e') inclusive electron scattering, and notes that the enhancement
appears only in the transverse channel.  The model is then applied by
scaling the magnetic form factors in the QE case.  This serves to
produce a prediction for the total cross section $\sigma(E)$.  But
because it is designed around a reweight of QE interactions, it can
not capture the complexity that the more detailed models do.
Certainly, such an implementation would poorly predict the energy
transfer and nucleon kinematics.

\subsection*{tuning options and uncertainties.}

This is an area for further exploration, especially as additional
modeling efforts described above mature.  As of this writing, we have not been so
creative about what is needed.   Early readers of this document might
suggest a few things to look at.

By construction, the Nieves model takes only one parameter from neutrino
data, the Axial vector mass (assuming a dipole), which historically
they have set to $M_A = 1.05$ GeV.  Likewise there is a treatment of
the Delta axial form factor.   In that sense, it is the only
thing available to be tuned.  Changing $M_A$ probably has
normalization effect more than anything else, with some mild $Q^2$
effect.  

Of course, experimentalists are more creative than that.
All current models listed above simplify something to allow calculations to proceed with
reasonable time, and neutrino-nucleus data sets still beg for
creative interpretation of models.  So variations between models, expressed as a
normalization or magnitude might be an okay way to start.  As the community
enhances our ability to compare models to neutrino-nucleus data, that
may suggest some additional targeted constraints.

Its conceivable to change the assumptions about where the model draws nucleons and what
it does with them.  For example, code up a blatantly asymmetric sharing of
energy and momentum transfer, or make the sharing artificially more
symmetric than the Lorentz boost.  As
it is, the model seems to produce as moderate a result as could be expected.

Finally, one could shave off near the momentum-transfer limit from 1.0 to 1.2 GeV, or enhance it.   This might be
especially important if someone's analysis is sensitive to the region
where the model is truncated.  The tail of the distribution of a
background process for some DIS study, for example.






\section*{Acknowledgements}

We are grateful for the assistance of the GENIE authors who
contributed significantly to this implementation.  The basic skeleton was
built off of Steve Dytman's ``empirical model'' code described in the
previous section,
and the two models still share significant code.
Gabe Perdue had
many helpful suggestions that improved this document, in addition to
being a vital resource to getting the code into GENIE.  The original
model authors Juan Nieves and Manuel Vicente Vacas and co-conspirator Federico
Sanchez had several technical suggestions
in the process from Ref.~\cite{Gran:2013kda} to now and comments on
this manuscript. 

We thank the GENIE team for the incubator workshop, hosted by
Fermilab in March 2013 where J.S. began this project, and for
Steve Dytman's continued support toward its completion. 
This work was supported by NSF awards 1306944 and 1607381 to the
University of Minnesota Duluth, and DOE grant number DE-FG02-93ER40788
to Colorado State University.  Also, J.S. was supported by the Colorado State
University Programs of Research and Scholarly Excellence.

\bibliography{mec-genie-note}

\end{document}